\documentclass{aa}  

\usepackage{graphicx}
\usepackage{txfonts}

\usepackage{natbib}
\bibpunct{(}{)}{;}{a}{}{,}
\begin{document} 

   \title{The orbits of subdwarf-B + main-sequence binaries}

   \subtitle{III. The period -- eccentricity distribution.}

   \author{J.~Vos
          \inst{1}
          \and
	  R.H.~\O{}stensen
	  \inst{2}
	  \and
	  M. Vuckovic
	  \inst{1}
          \and
          H. Van Winckel
	  \inst{3}
          }
   
   \institute{ Instituto de F\'{\i}sica y Astronom\'{\i}a, Universidad de Valparaiso, Gran Breta\~{n}a 1111, Playa Ancha, Valpara\'{\i}so 2360102, Chile\\
               \email{joris.vos@uv.cl}
               \and
               Department of Physics, Astronomy, and Materials Science, Missouri State University, Springfield, MO 65804, USA
               \and
               Instituut voor Sterrenkunde, KU Leuven, Celestijnenlaan 200D, B-3001 Leuven, Belgium\\
              }

   \date{Received \today; accepted ???}

 
  \abstract
   {The predicted orbital-period distribution of the subdwarf-B (sdB) population is bi-modal with a peak at short ( $<$ 10 days) and long ( $>$ 500 days) periods. Observationally,  many short-period sdB systems are known, but only few wide sdB binaries have been studied in detail. Based on a long-term monitoring program the wide sdB sample has been increased, discovering an unexpected positive correlation between the eccentricity and orbital period.} 
   {In this article we present the orbital solution and spectral analysis of four new systems, BD$-$7$^{\circ}$5977, EC\,11031--1348, TYC\,2084--448--1 and TYC\,3871--835--1, and update the orbital solution of PG\,1104+243. Using the whole sample of wide sdB binaries, we aim at finding possible correlations between orbital and spectral properties. The ulitmate goal is to improve theoretical models of Roche-lobe overflow (RLOF).}
   {High-resolution spectroscopic time series were obtained to determine the radial velocities of both the sdB and MS components. Literature photometry was used to construct the spectral-energy distributions, which were fitted with atmosphere models to determine the surface gravities and temperatures of both components in all systems. Spectral parameters of the cool companion were verified using the GSSP code. Furthermore the amount of accreted mass was estimated.}
   {Orbital parameters matching the earlier observed period-eccentricity relation were found for three systems, while TYC\,2084--448--1 is found to have a lower eccentricity than expected from the period-eccentricity trend indicated by the other systems. Based on new observations, the orbit of PG\,1104+243 has a small but significant eccentricity of 0.04 $\pm$ 0.02, matching that of the other systems with similar periods. Furthermore, a positive correlation between accreted mass and orbital period was found, as well as a possible relation between the initial mass-ratio and the final period-eccentricity.}
   {The wide sdB-binary sample shows interesting possible correlations between orbital and spectral properties. However, a larger sample is necessary to statistically validate them.}


   \maketitle
%

\defcitealias{Vos2012}{Paper I}
\defcitealias{Vos2013}{Paper II}

\newcommand\kms{km s$^{-1}$}
\newcommand\vsini{v$_{\rm r}\sin{i}$}

\section{Introduction}\label{s:instroduction}
Hot subdwarf-B (sdB) stars are core-helium-burning stars with a very thin hydrogen envelope (M$_{\rm{H}}$ $<$ 0.02 $M_{\odot}$), and a mass close to the core-helium-flash mass $\sim$ 0.47 $M_{\odot}$ \citep{Saffer1994, Brassard2001}. These hot subdwarfs are the main source for the UV-upturn in early-type galaxies \citep{Green1986, Greggio1990, Brown1997}, and are found in all galactic populations. Furthermore, \citet{Heber1998} discovered that the chemical peculiarities in the photospheres of sdB stars, including the strong He-depletion, are caused by diffusion processes.

To ignite He-burning in the core while at the same time not have sufficient hydrogen remaining to sustain H-shell burning, the sdB progenitor needs to lose most of its hydrogen envelope near the tip of the red-giant branch (RGB). A variety of different possible formation channels has been proposed, but the current consensus is that sdB stars can only be formed through binary interaction. The main binary formation channels thought to contribute to the sdB population are the common-envelope (CE) ejection channel \citep{Paczynski1976, Han2002}, the stable Roche-lobe overflow (RLOF) channel \citep{Han2000, Han2002}, and the formation of an single sdB star as the end product of a binary white-dwarf (WD) merger \citep{Webbink1984}. 

Based on binary-population-synthesis (BPS) studies including the CE-ejection channel, the stable-RLOF channel and the WD merger scenario, \citet{Han2002, Han2003} and \citet{Chen2013} found that the CE-ejection channel produces close binaries with periods on the order of hours to a few days. The RLOF channel produces wide sdB binaries with periods up to $\sim$1600\ d, and the WD-merger channel leads to single sdB stars with a potentially higher mass. A detailed review of hot subdwarf stars is given by \citet{Heber2016}. 

The short-period sdB binaries have been the focus of many observational studies, and more then 150 of these systems are currently known \citep[][and references therein]{Kawka2015, Kupfer2015}. The observed properties of these post-CE sdB binaries correspond well with the results of BPS studies. However, the first observational results of wide sdB binaries based on long term observing campains have only recently been published \citep{Deca2012, Vos2012, Barlow2013, Vos2013}. 

This article presents the results of the long-term observing program with the Mercator telescope started in 2009 \citep{Gorlova2014}. Four sdB binaries included in the program have solved orbits, and were presented in \citep[][\citetalias{Vos2012}]{Vos2012} and \citep[][\citetalias{Vos2013}]{Vos2013}. Here, we present the orbital solution, and a spectral analysis for the two systems of which preliminary orbits have been published (BD$-$7$^{\circ}$5977 and TYC\,3871--835--1, \citealt{Vos2014}), and two previously unpublished systems (EC\,11031--1348 and TYC\,2084--448--1). Furthermore, the orbital solution of PG\,1104+243 is refined based on the latest spectra. The orbital solution of these five systems is presented in Sect\,\ref{s:spectroscopy}. The spectral energy density (SED) of the four systems, and the derived atmospheric parameters are given in Sect\,\ref{s:SED}, while the atmospheric parameters determined from fitting the spectra using the grid search in stellar parameters (GSSP) package are discussed in Sect\,\ref{s:gssp}. 

The only other wide sdB binaries published in the literature with solved orbits are PG\,1018-047 \citep{Deca2012, Deca2017}, PG\,1449+653 and PG\,1701+359 \citep{Barlow2013}. Together with the eight systems presented in this series, the total wide sdB binary sample consists of eleven systems. An overview of the main parameters of these systems is given in Sect\,\ref{s:wide_sdB_sample}, based on which an estimate of the mass accreted by the cool companion during the RLOF phase of the sdB progenitor is made. In Sect\,\ref{s:period_ecc_distribution} the period-eccentricity distribution of all currently known wide sdB-binaries is discussed.

\section{Spectroscopy}\label{s:spectroscopy}

High resolution spectroscopic observations of BD$-$7$^{\circ}$5977, EC\,11031--1348, TYC\,2084--448--1, TYC\,3871--835--1 and PG\,1104+243 were obtained with the High Efficiency and Resolution Mercator Echelle Spectrograph (HERMES, R = 85\,000, 3770-9000 \AA, \citealt{Raskin2011}) spectrograph at the 1.2 Mercator telescope at the Roque de los Muchachos Observatory, La Palma. The observing method is unchanged from \citetalias{Vos2012} and \citetalias{Vos2013}. The high-resolution mode of HERMES was used, and Th-Ar-Ne exposures were made at the beginning, middle and end of the night. The exposure time was calculated to reach a signal-to-noise ratio (S/N) of 25 or higher in the $V$--band. The reduction of the spectra was performed using the 5th version of the HERMES pipeline, which includes the barycentric correction. 

In total there are 45 spectra of BD$-$7$^{\circ}$5977, 52 spectra of  EC\,11031--1348, 51 spectra of TYC\,2084--448--1, 57 spectra of TYC\,3871--835--1 and 99 spectra of PG\,1104+243. All observations were taken between June 2009 and September 2016.

%

\subsection{Radial velocities}
The determination of the radial velocities of the MS component is straightforward as they have many clear metal lines visible in the spectra. Even with a low S/N, accurate velocities can be derived. For the systems with a slow rotating MS component, the cross-correlation (CC) method of the HERMES pipeline, based on a discrete number of line positions, is used. The sdB component only has a few H and He lines, which are avoided in this cross-correlation. The CC function is used on orders 55-74 (4780 - 6530 \AA) as these orders give the best compromise between maximum S/N and absence of telluric lines. The errors on the radial velocities are calculated taking into account the formal errors on the Gaussian fit to the normalised CC function and the error due to the stability of the wavelength calibration. See \citetalias[Sect.\,2]{Vos2012} for more details of this procedure.

In the case of a fast rotating MS component, a cross-correlation with a template spectrum is used. This template spectrum is calculated based on an estimate of the spectral parameters for these systems, and in a second iteration based on the spectral parameters determined for these systems using the GSSP code and the SED fitting approach (see Sect.\,\ref{s:gssp}). The cross correlation is preformed on the same wavelength range (4780 - 6530 \AA), and the errors are determined by using a Monte Carlo (MC) approach. Gaussian noise is added to the spectrum, after which the CC is repeated, and the error is determined by the standard deviation of the different RVs calculated in 250 iterations. 

Determining the RVs of the sdB components is more challenging as the only non-contaminated line is the \ion{He}{i} blend at 5875.61 \AA. To increase the S/N, spectra taken with only a few days in between are co-added together. The period for which spectra are co-added depends on the system, but does not exceed 50 days. This is equivalent to 5\,\% of the orbital period, thus the amount of smearing or broadening of the He line is small. To obtain the RVs, the \ion{He}{I} line is cross correlated with a high-resolution synthetic sdB spectrum (T$_{\rm eff}$ = 30000 K, $\log{g}$ = 5.50 dex) from the LTE grids of \citet{Heber2000} , matched to the resolution of the HERMES spectra. The error is determined using the same MC method used for the fast rotating MS components. A more elaborate explanation of the RV calculations is given in \citetalias[Sect.\,2]{Vos2012}.

\begin{figure*}
    \includegraphics{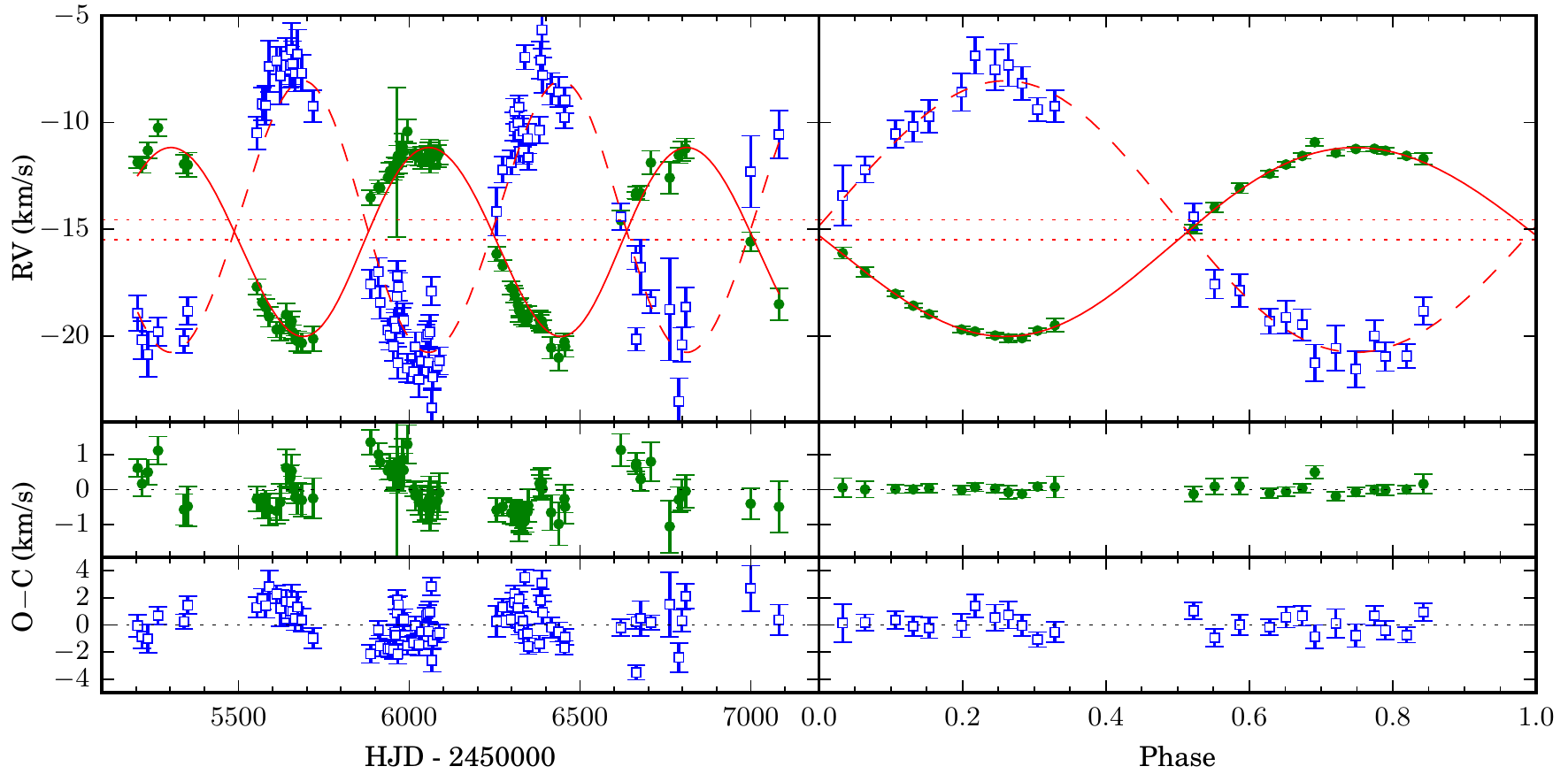}
    \caption{The radial velocity curve and residuals (O$-$C) of PG\,1104+243. The radial velocities of the cool companion are plotted in green filled circles, while those of the sdB are shown in open blue squares. The best fitting Keplerian orbit is shown as a red full line for the cool companion and red dashed line for the sdB. Left: the radial velocities of the individual spectra. Right: the radial velocities derived for the phase binned spectra.}
    \label{fig:rv_curves_pg1104}
\end{figure*}

The exact procedure used for the individual systems is described in Sect.\,\ref{s:rv-results}. The RVs for both the MS and sdB components are given in the tables in Appendix\,\ref{a:rv_tables}, and are plotted in Figs.\,\ref{fig:rv_curves} and \ref{fig:rv_curves_pg1104}.

\begin{figure*}
    \includegraphics{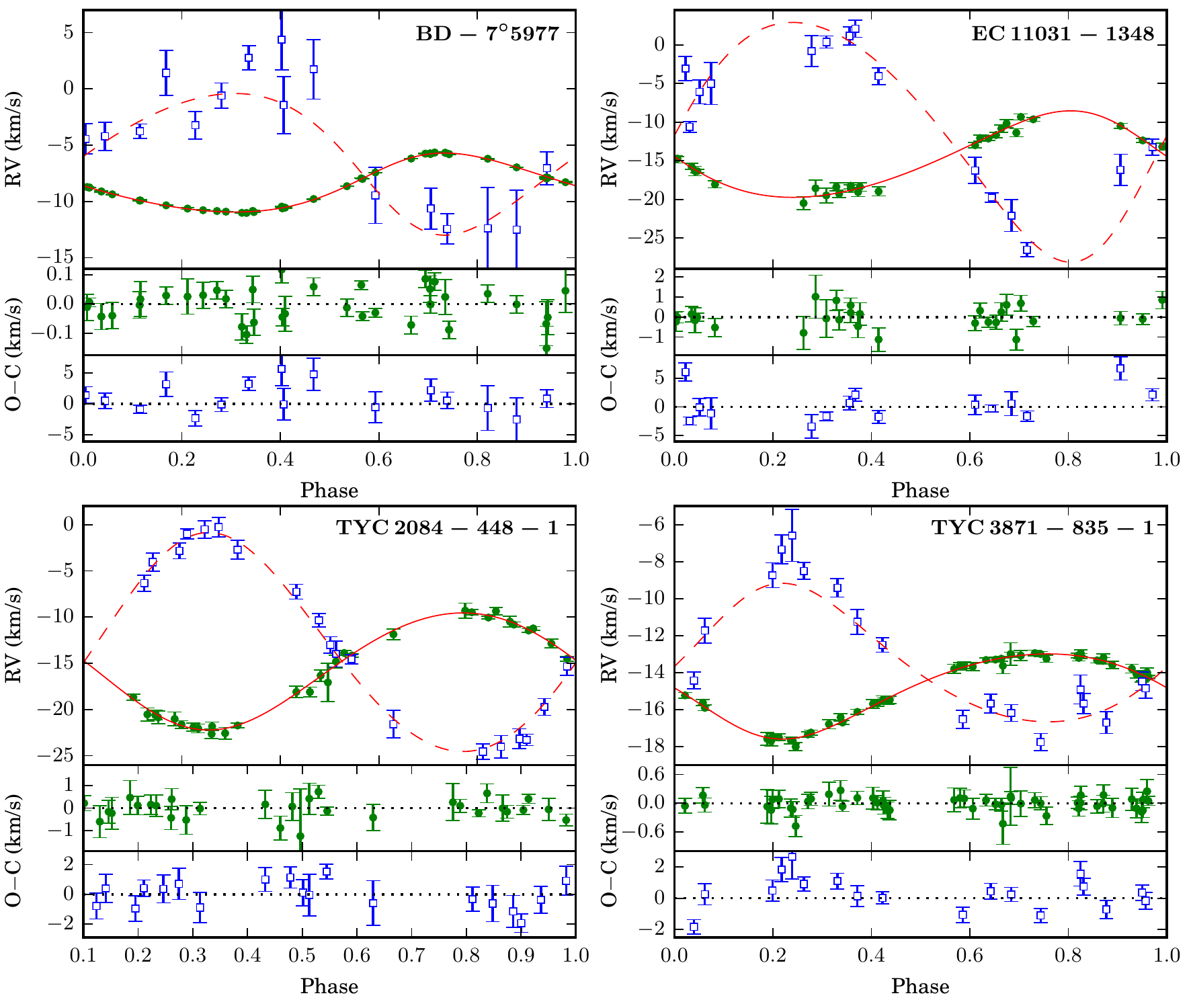}
    \caption{The radial velocity curves and residuals (O$-$C) of BD$-$7$^{\circ}$5977, EC\,11031--1348, TYC\,2084--448--1 and TYC\,3871--835--1. The radial velocities of the cool companion are plotted in green filled circles, while those of the sdB are shown in open blue squares. The best fitting Keplerian orbit is shown in red full line for the cool companion and red dashed line for the sdB.}
    \label{fig:rv_curves}
\end{figure*}

\subsection{Orbital parameters} \label{s:orbital_param}
To calculate the orbital parameters of the sdB and MS components we follow the same method as described in \citetalias{Vos2012} and \citetalias{Vos2013}. Fitting a Keplerian orbit with eight free parameters, the orbital period ($P$), time of periastron ($T_0$), eccentricity ($e$), angle of periastron ($\omega$), and for both components their amplitudes ($K_{\rm{MS}}$ and $K_{\rm{sdB}}$) and systemic velocities ($\gamma_{\rm{MS}}$ and $\gamma_{\rm{sdB}}$), to the radial velocity measurements. The resulting parameters and their errors are given in Table\,\ref{tb:orbital_parameters}, and the best fitting Keplerian orbits are shown in Figs.\,\ref{fig:rv_curves} and \ref{fig:rv_curves_pg1104}.


\subsection{Results}\label{s:rv-results}
The method to derive the radial velocities from the HERMES spectra and the calculated orbital parameters for each system are discussed below.

\begin{description}
 \item[BD$-$7$^{\circ}$5977:] The companion in this system is a subgiant (SG), which dominates the spectrum at all observed wavelengths. The spectral lines of the SG are sharp, thus the HERMES CC function was used with an Arcturus template. The average RV error is 0.003 \kms\ which is due to the many sharp lines of the SG. To determine the RVs of the sdB, the spectra taken within a period of 50 days (4 \% of the orbital period) are co-added to increase the S/N. The \ion{He}{I} line at 5875.61 \AA\ shows some contribution of the SG. To diminish this effect, a synthetic spectrum of the SG component around the \ion{He}{I} line, shifted to the velocity of the companion, is subtracted from the observed spectra before the cross correlation. The average error on the RVs of the sdB is 2 \kms. \\ 
 BD$-$7$^{\circ}$5977 has an orbital period of 1262 $\pm$ 1 days and an eccentricity of 0.16 $\pm$ 0.01. Due to the small errors on the RVs of the SG component, the period and eccentricity can be determined with high accuracy. The large errors on the RVs of the sdB component result in a high uncertainty on the derived amplitude and mass ratio ($q$ = 0.4 $\pm$ 0.1). The average O-C for the SG companion is $\sigma_{\rm SG} = 0.04$ \kms, while for the sdB it is much higher at $\sigma_{\rm sdB} = 1.8$ \kms.\\
 
 \item[EC\,11031--1348:] With a V-magnitude of 11.5, this system is faint for HERMES, and the MS component is a fast rotator. To determine the RVs of the MS component, all spectra taken within 10 days are co-added to increase the S/N, and a cross correlation with a rotationally broadened synthetic spectrum was used.  The average error for the MS component is 0.28 \kms. The \ion{He}{I} line of the sdB component is weak, but is not contaminated by any lines from the companion. To increase the S/N, spectra within 50 days are co-added. The average RV error for the sdB component is 1.4 \kms.\\
 The derived orbital period is 1099 $\pm$ 6 days and it has an eccentricity of 0.17 $\pm$ 0.03. Both the spread of the MS RVs, $\sigma_{\rm MS} = 0.4$ \kms, and the sdB RVs, $\sigma_{\rm sdB} = 2.0$ \kms, are the largest compared to the other three systems.\\
 
 \item[TYC\,2084--448--1:] This system also contains a fast rotating MS companion. Spectra taken within 20 days were co-added, and a cross correlation with a rotationally broadened synthetic spectrum was used to determine the MS RVs. The average RV error for the MS component is 0.10 \kms. To determine the RVs of the sdB, all spectra within 50 days were co-added. The average sdB RV error is 0.78 \kms.\\
 The period of this system is very close to that of EC\,11031--1348, $P = 1098 \pm 5$ days, but the eccentricity is much lower, $e = 0.05 \pm 0.03$. The eccentricity test shows that the chance of this orbit to be circular is less than 1\%. The average O--C for the MS and sdB components are respectively $\sigma_{\rm MS} = 0.3$ \kms and $\sigma_{\rm sdB} = 0.8$ \kms.\\
 
 \item[TYC\,3871--835--1:] The companion is a slowly rotating solar like star, thus the HERMES CC function was used on the original spectra, with a G2 type template. The average RV error for the MS component is 0.05 \kms. For the sdB component, spectra within 50 days are co-added, resulting in an average sdB RV error of 0.60 \kms.\\
 This system has one of the longest orbital periods of 1263 $\pm$ 5 days, and an eccentricity of 0.16 $\pm$ 0.02. Both the MS and sdB RVs fit the Keplerian curves well with an average O--C of $\sigma_{\rm MS} = 0.1$ \kms and $\sigma_{\rm sdB} = 0.8$ \kms.\\
 
 \item[PG\,1104+243:] This system has already been analysed in \citetalias{Vos2012}. The companion is a solar type star which shows small but significant deviations from the periodic signal in the RV curves. As this star is not in any known instability strip of the HR diagram, but is very close to the sun in both effective temperature and surface gravity, the variations were attributed to spots. With the new spectra that cover multiple orbits, the effects of spots can be minimised by merging spectra together based on orbital phase. The orbital period of this system is well known as there are RV measurements available from earlier epochs. Based on this orbital period, the spectra are co-added in phase bins of 0.02, resulting in 25 phase added spectra. The RVs of the MS component of these spectra are determine using the HERMES CC function with a G2 type template, and their average error is 0.032 \kms. The calculation of the sdB RVs is done on the same phase binned spectra. The average sdB RV error is 0.78 \kms. In Fig.\,\ref{fig:rv_curves_pg1104}, the original and phased-added RVs are shown next to each other, together with the best fitting Keplerian curve. The spread in the residuals is significantly smaller for the phase-added spectra.\\
 Most of the orbital parameters have not changed with respect to those presented in \citetalias{Vos2012}, but with the new RVs, the orbit is found to have a small eccentricity of 0.04 $\pm$ 0.02. The eccentricity test indicates that the chance of a circular orbit is less then 1\%. The O--C values for the original measurements are $\sigma_{\rm MS} = 0.5$ \kms and $\sigma_{\rm sdB} = 1.2$ \kms, while those for the phase-added spectra are $\sigma_{\rm MS} = 0.08$ \kms and $\sigma_{\rm sdB} = 0.5$ \kms.\\
\end{description}

\begin{table*}
 \centering
 \caption{Spectroscopic orbital solutions for both the main-sequence (MS) and subdwarf-B (sdB) component of BD$-$7$^{\circ}$5977,  EC\,11031--1348, TYC\,2084--448--1, TYC\,3871--835--1 and PG\,1104+243. The parameters are: orbital period ($P$), time of periastron ($T_0$), eccentricity ($e$), angle of periastron ($\omega$), mass ratio ($q$) and for both component the amplitude ($K$), system velocity ($\gamma$), reduced mass ($M\sin^3{i}$) and the reduced semi-major axis ($a \sim{i}$).}
 \label{tb:orbital_parameters}
 \begin{tabular}{lr@{ $\pm$ }lr@{ $\pm$ }lr@{ $\pm$ }lr@{ $\pm$ }lr@{ $\pm$ }l}
 \hline\hline
 \noalign{\smallskip}
 Parameter    &   \multicolumn{2}{c}{BD$-$7$^{\circ}$5977}   &   \multicolumn{2}{c}{EC\,11031--1348}   &   \multicolumn{2}{c}{TYC\,2084--448--1}   &   \multicolumn{2}{c}{TYC\,3871--835--1}   &   \multicolumn{2}{c}{PG\,1104+243}   \\\hline
 \noalign{\smallskip}
$P$ (d)                              &  1262     &  1     &  1099     &  6     &  1098     &  5     &  1263     &  5     &  755       &  3     \\
$T_0$ (d)                            &  2454971  &  4     &  2456600  &  20    &  2456054  &  58    &  2454075  &  18    &  2450480   &  8     \\
$e$                                  &  0.16     &  0.01  &  0.17     &  0.03  &  0.05     &  0.03  &  0.16     &  0.02  &  0.04      &  0.02  \\
$\omega$                             &  5.5      &  0.1   &  3.9      &  0.1   &  5.63     &  0.33  &  2.83     &  0.08  &  0.70      &  0.02  \\
$q$ (sdB/MS)                         &  0.4      &  0.1   &  0.36     &  0.02  &  0.51     &  0.02  &  0.54     &  0.05  &  0.70      &  0.02  \\\noalign{\smallskip}
$K_{\rm MS}$ (\kms)                  &  2.62     &  0.01  &  5.55     &  0.15  &  6.30     &  0.10  &  2.31     &  0.04  &  4.42      &  0.04  \\
$\gamma_{\rm MS}$ (\kms)             &  -8.62    &  0.00  &  -14.75   &  0.08  &  -15.56   &  0.06  &  -14.98   &  0.02  &  -15.59    &  0.06  \\
M$_{\rm MS} \sin^3{i}$ (M$_{\odot}$) &  0.06     &  0.03  &  0.77     &  0.07  &  0.48     &  0.03  &  0.02     &  0.01  &  0.057     &  0.002 \\
a$_{\rm MS} \sin{i}$ (R$_{\odot}$)   &  64       &  15    &  120      &  6     &  136      &  5     &  56       &  3     &  66        &  1     \\\noalign{\smallskip}
$K_{\rm sdB}$ (\kms)                 &  6.2      &  1.0   &  15.5     &  0.6   &  12.25    &  0.30  &  4.24     &  0.20  &  6.34      &  0.2   \\
$\gamma_{\rm sdB}$ (\kms)            &  -5.5     &  1.0   &  -11.4    &  0.3   &  -13.30   &  0.20  &  -13.36   &  0.12  &  -13.8     &  0.2   \\
M$_{\rm sdB} \sin^3{i}$ (M$_{\odot}$)&  0.03     &  0.01  &  0.28     &  0.04  &  0.25     &  0.02  &  0.010    &  0.005 &  0.039     &  0.004 \\
a$_{\rm sdB} \sin{i}$ (R$_{\odot}$)  &  152      &  25    &  330      &  12    &  330      &  11    &  104      &  8     &  95        &  2     \\
\hline
 \end{tabular}
\end{table*}

\section{Spectral Energy Distribution}\label{s:SED}
The spectral-energy distribution (SED) of the systems can be used to determine the effective temperature and surface gravity of both the MS and sdB component. This is done by fitting model SEDs to the photometric SEDs. 

\subsection{Photometry}
We collect photometry for all four systems in the sdB database\footnote{http://catserver.ing.iac.es/sddb/} \citep{Oestensen2006}, which containt a compilation of data on hot sdB stars from the literature. These photometric measurements are supplemented with photometry obtained from the AAVSO Photometric All-Sky Survey (APASS) DR9 \citep{Henden2016} and 2MASS \citep{Skrutskie2006}. The obtained photometry for each target is listed in Appendix\,\ref{tb:BD-7_phot} to \ref{tb:TYC3871_phot}. Accurate photometric measurements at both short and long wavelengths are used to establish the contribution of the hot sdB component and the cool MS component.

\begin{figure*}
    \includegraphics{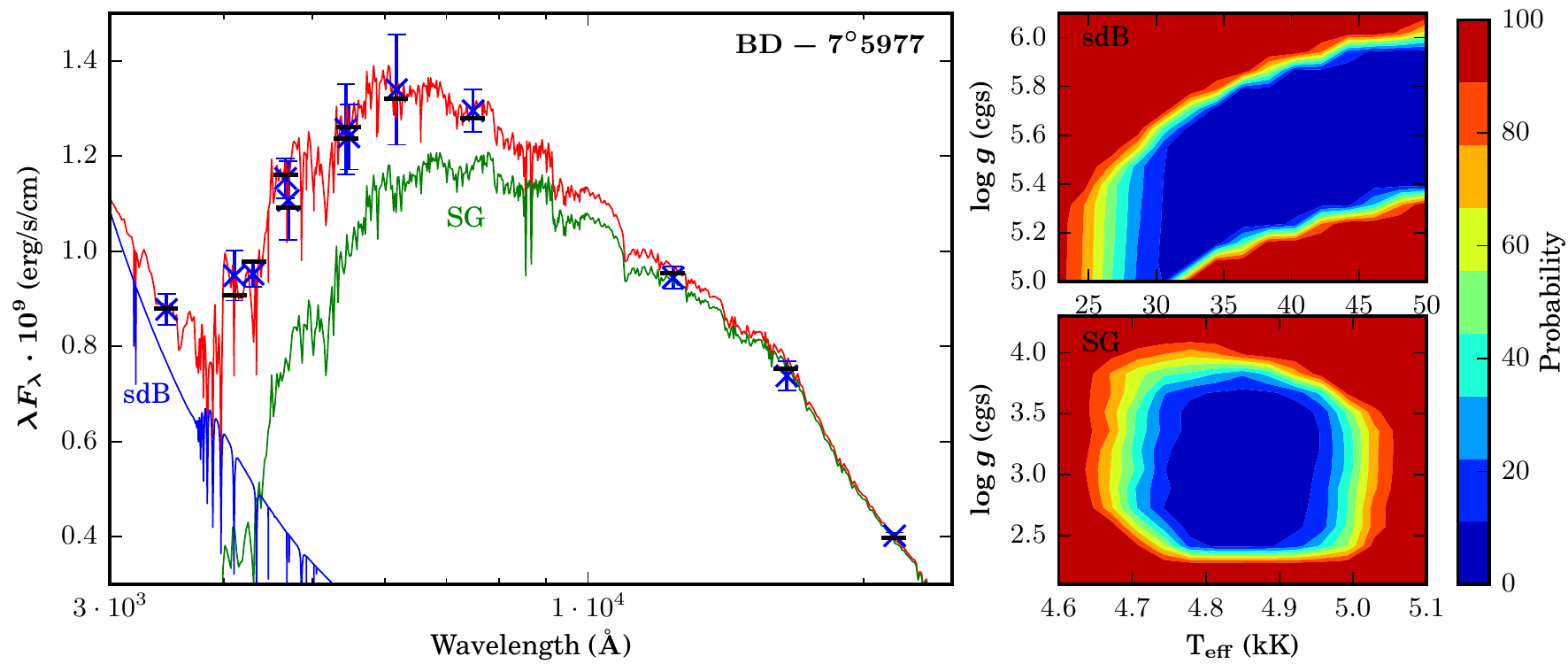}
    \caption{The photometric SED of BD-7$^{\circ}$5977 with the best fitting model (left panel) and the confidence intervals for the effective temperature and surface gravity of both components (right panels). The observed photometry is plotted in blue crosses, while the synthetic best fit photometry is plotted in black horizontal line. The best fitting binary model is shown in a red line, while the models for the cool companion and sdB star are show respectively in green line and blue line.}
    \label{fig:BD-7_SEDfit}
\end{figure*}

\begin{figure*}
    \includegraphics{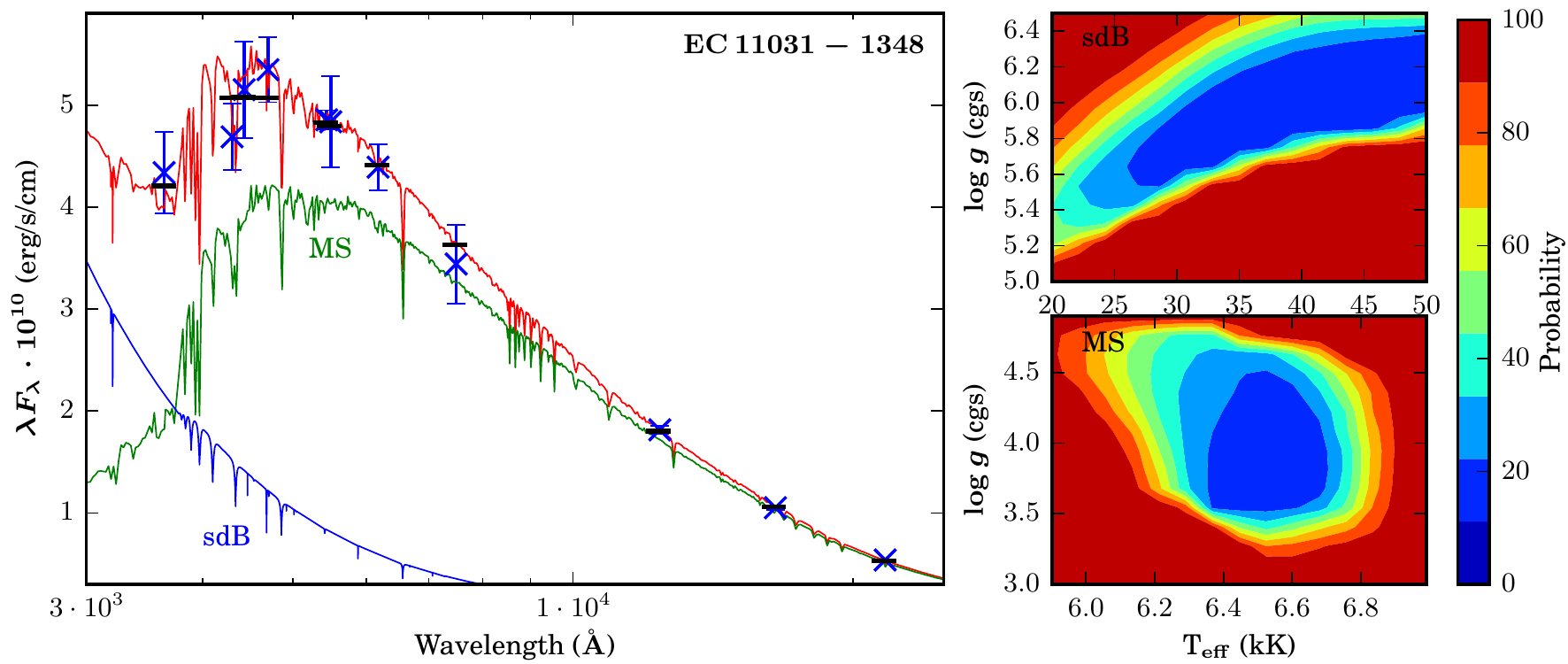}
    \caption{Same as Fig.\,\ref{fig:BD-7_SEDfit} but for EC\,11031--1348.}
    \label{fig:EC11031_SEDfit}
\end{figure*}

\begin{figure*}
    \includegraphics{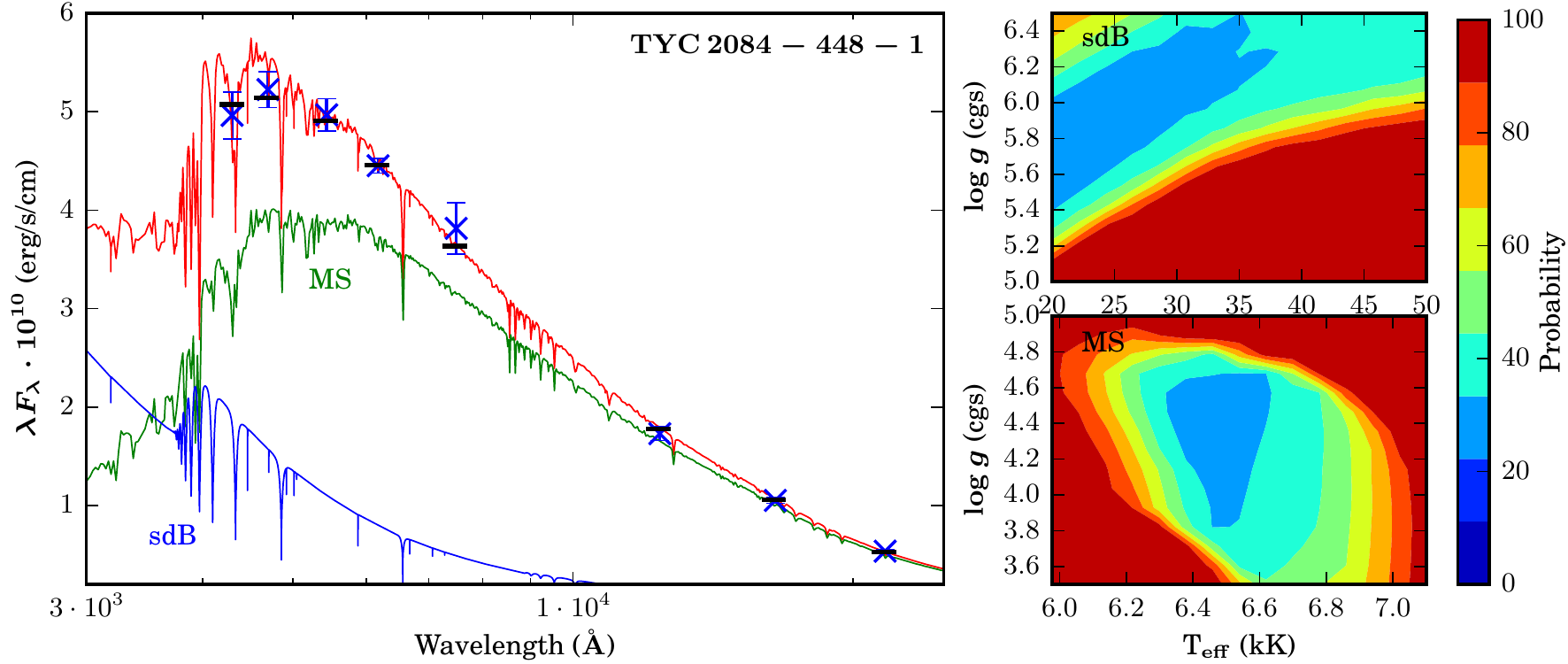}
    \caption{Same as Fig.\,\ref{fig:BD-7_SEDfit} but for TYC\,2084--448--1.}
    \label{fig:TYC2084_SEDfit}
\end{figure*}

\begin{figure*}
    \includegraphics{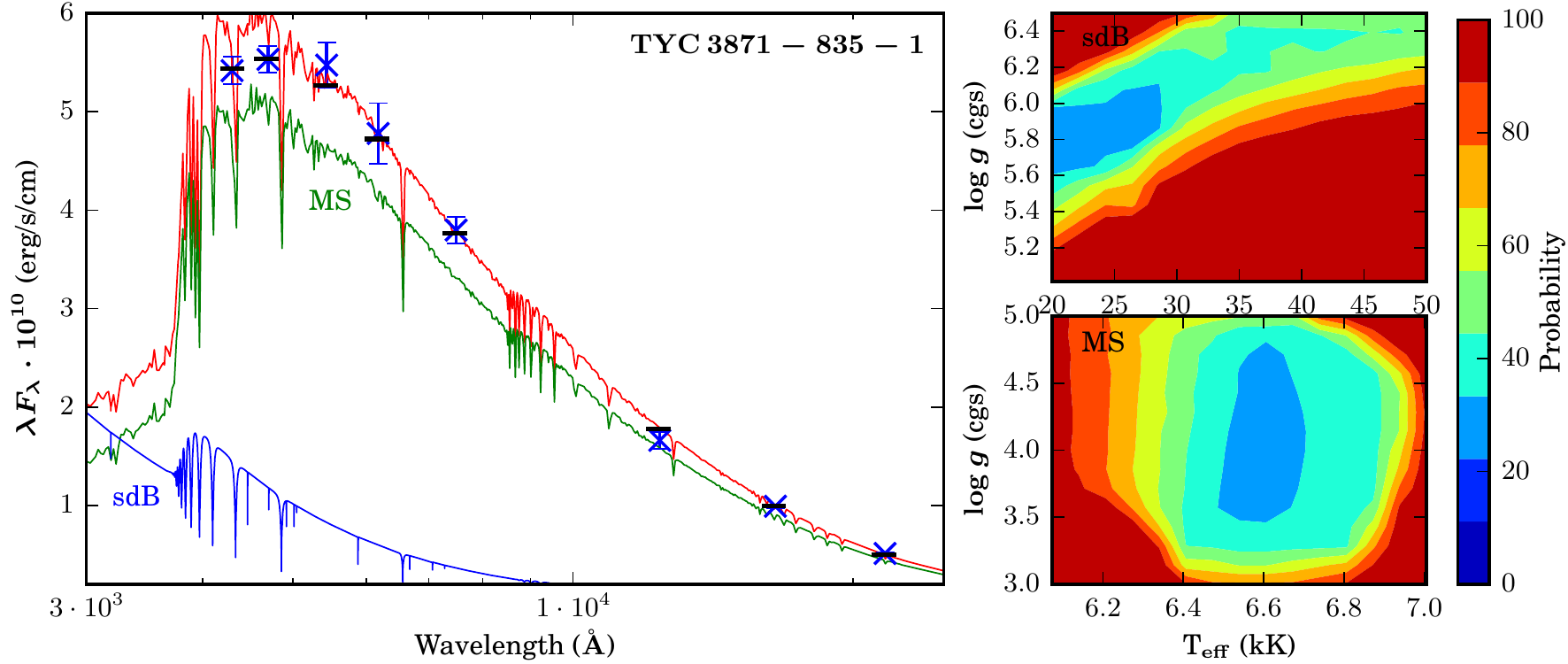}
    \caption{Same as Fig.\,\ref{fig:BD-7_SEDfit} but for TYC\,3871--835--1.}
    \label{fig:TYC3871_SEDfit}
\end{figure*}

\subsection{SED fitting}
The SED fitting method is the same as described in \citetalias[Sect.\,3]{Vos2012} and \citetalias[Sect.\,3.2]{Vos2013}. The observed photometry is fitted with a synthetic SED integrated from model atmospheres. For the MS component Kurucz atmosphere models \citep{Kurucz1979} ranging in effective temperature from 3000 to 9000 K, and in surface gravity from $\log{g}$=2.0 dex (cgs) to 5.0 dex (cgs) are used. For the hot sdB component TMAP (T\"{u}bingen NLTE Model-Atmosphere Package, \citealt{Werner2003}) atmosphere models with a temperature range from 20000 K to 50000 K, and $\log{g}$ from 5.0 dex (cgs) to 6.5 dex (cgs) are used.

To fit the model SEDs to the photometric observations, we use the two step method described in \citetalias{Vos2012} and \citetalias{Vos2013}. This method consists of a grid based fitting approach described in \citet{Degroote2011}, where the entire parameter space is scanned by randomly picking 5\,000\,000 points in the parameter space and calculating the $\chi^2$. The 50 grid points with the lowest $\chi^2$ are used as starting point for a least-squares minimizer which determines the final result.

In total the SED fit needs to consider eight parameters: the effective temperatures ($T_{\rm{eff,MS}}$ and $T_{\rm{eff,sdB}}$), surface gravities ($g_{\rm{MS}}$ and $g_{\rm{sdB}}$) and radii ($R_{\rm{MS}}$ and $R_{\rm{sdB}}$) of both components, the interstellar reddening $E(B-V)$ and the distance ($d$) to the system. The $E(B-V)$ of both components is considered equal. The distance to the system is calculated by scaling the synthetic models to the photometric observations. Because the mass-ratio of the binary is known, the radii of both components can be coupled to their surface gravity, reducing the number of free parameters to six (\citetalias[Sect.\,3]{Vos2012}).


\subsection{Results}\label{s:SED-results}
The SED fit for PG\,1104+243 is described in \citet{Vos2012}. As the mass ratio did not change with respect to the radial velocity curve given in that article, the SED results will remain the same as well. The resulting spectral parameters for the other four systems are given in Table\,\ref{tb:sed_gssp_parameters}, while the individual fits are shown in Figs.\,\ref{fig:BD-7_SEDfit}, \ref{fig:EC11031_SEDfit}, \ref{fig:TYC2084_SEDfit} and \ref{fig:TYC3871_SEDfit}. The SED fits for each system are discussed below.

\begin{table*}
  \centering
 \caption{Spectral parameters derived from the SED fits and the GSSP analysis of BD$-$7$^{\circ}$5977,  EC\,11031--1348, TYC\,2084--448--1 and TYC\,3871--835--1. The errors of the parameters derived with GSSP are 1 $\sigma$ errors. The dilution factor given for the SED fit is the dilution of the cool companion calculated in wavelength range 5800--6500\AA, and is included to provide easy comparison with the GSSP dilution factor. }
 \label{tb:sed_gssp_parameters}
 \begin{tabular}{lr@{ $\pm$ }lr@{ $\pm$ }lr@{ $\pm$ }lr@{ $\pm$ }l}
 \hline\hline
 \noalign{\smallskip}
 Parameter    &   \multicolumn{2}{c}{BD$-$7$^{\circ}$5977}   &   \multicolumn{2}{c}{EC\,11031--1348}   &   \multicolumn{2}{c}{TYC\,2084--448--1}   &   \multicolumn{2}{c}{TYC\,3871--835--1}  \\\hline
 \noalign{\smallskip}
 \multicolumn{9}{c}{SED analysis}\\ \noalign{\smallskip}
T$_{\rm eff,MS}$ (K)      &  4850  &  250   &  6600  &  400   &  6450  &  600   &  6500  &  500  \\
$\log{g}_{\rm MS}$ (dex)  &  2.80  &  0.40  &  4.00  &  0.50  &  4.40  &  0.60  &  4.30  &  0.70 \\
T$_{\rm eff,sdB}$ (K)     &  30000 &  7000  &  28000 &  5000  &  25000 &  7000  &  24000 &  7000 \\
$\log{g}_{\rm sdB}$ (dex) &  5.20  &  0.60  &  5.60  &  0.50  &  5.70  &  0.70  &  5.80  &  0.70 \\
Dilution                  &  0.84  &  0.15  &  0.86  &  0.10  &  0.82  &  0.12  &  0.79  &  0.14 \\
\hline\noalign{\smallskip}
\multicolumn{9}{c}{GSSP analysis}\\ \noalign{\smallskip}
T$_{\rm eff}$ (K)    &  4800  &  100   &  6400  &  250   &  6300  &  200   &  6200  &  150  \\
$\log{g}$ (dex)      &  3.00  &  0.40  &  4.30  &  0.50  &  4.50  &  0.40  &  4.40  &  0.30 \\
$[$Fe/H$]$ (dex)     &  -0.20 &  0.10  &  0.20  &  0.20  &  -0.10 &  0.10  &  0.10  &  0.10 \\
\vsini\ (\kms)       &  9     &  1     &  70    &  4     &  51    &  2     &  15    &  1    \\
Dilution             &  0.72  &  0.05  &  0.84  &  0.10  &  0.75  &  0.05  &  0.65  &  0.02 \\
\hline
 \end{tabular}
\end{table*}

\begin{description}
 \item[BD$-$7$^{\circ}$5977:] The subgiant companion is clearly visible in the SED, as it dominates the SED up until the U band. Next to the photometry from APASS and 2MASS, also Str\"{o}mgren photometry from the Space lab-1 Very Wide Field Survey of UV-excess objects \citep{Viton1991} is available for this object. There is a GAIA parallax available for BD$-$7$^{\circ}$5977, which could be used to fix the distance when fitting the SED. However, because this is a binary system, with an orbital period longer then the GAIA observing time frame, the orientation of the centre of light would change during the parallax observations. This effect would first have to be studied and quantified before an accurate distance can be derived for binary systems from the GAIA parallax. We aim at doing this in a future article.\\
 For the SG companion we find an effective temperature of 4850 $\pm$ 250 K and a surface gravity of 2.80 $\pm$ 0.40 dex. The confidence interval for the SG component is roughly following a Gaussian distribution. However, the parameters of the sdB component are less well constrained, due to the majority of the light in the fitted region originating from the SG component. This is reflected in the confidence intervals, which show a large spread of equally possible spectral parameters. The best fitting being, T$_{\rm eff, sdB}$ = 30\,000 $\pm$ 7000 K and $\log{g}_{\rm sdB} = 5.20 \pm 0.60$ dex. \\
 
 \item[EC\,11031--1348:] Together with the APASS and 2MASS photometry also Johnson U, B and V measurements from the Edinburgh-Cape Blue Object survey of \citet{Kilkenny1997} are available. The uncertainties on the APASS photometry are larger than for the APASS photometry of the other three systems. The addition of the Johnson photometry helps making up for the larger errors. The SED fit results in an effective temperature of 28\,000 $\pm$ 5000 K and 6600 $\pm$ 400 K for respectively the sdB and the cool companion. The surface gravities found are $\log{g}_{\rm sdB} = 5.60 \pm 0.50$ dex and $\log{g}_{\rm MS} = 4.0 \pm 0.50$ dex respectively. The best fitting surface gravity of 4.0 dex for the MS component is low for an F5-type star, but the errors are large enough to allow for a higher $\log{g}$.\\
 
 \item[TYC\,2084--448--1:] Only APASS and 2MASS photometry is available for this system. The SED fit shows T$_{\rm eff}$ = 24\,000 $\pm$ 6000 K and $\log{g} = 5.70 \pm 0.60$ dex for the sdB, while the F-type companion has an effective temperature of 6450 $\pm$ 600 K and a surface gravity of 4.40 $\pm$ 0.60 dex. The lack of available U-band photometry somewhat increases the confidence intervals for the sdB component. This is also noticeable in the case of TYC\,3871--835--1. \\
 
 \item[TYC\,3871--835--1:] As with TYC\,2084--448--1, only APASS and 2MASS photometry could be found for this system, and U-band photometry is lacking. Again an F-type companion is found with T$_{\rm eff}$ = 6500 $\pm$ 500 K and $\log{g} = 4.30 \pm 0.70$ dex, while the sdB component has T$_{\rm eff}$ = 25\,000 $\pm$ 5000 K and $\log{g} = 5.80 \pm 0.70$ dex.\\
\end{description}

With the exception of the SG component in BD$-$7$^{\circ}$5977, the uncertainties on the parameters derived from the SED fits are large. In the current sed fitting algorithm, the models are scaled to the observations by changing the distance. However, as GAIA parallaxes become available, and a correct distance can be derived also in the case of long-period binaries, the added constraints on the distance will significantly reduce the uncertainties on the SED parameters by forcing the luminosity of the model to match the observed luminosity and the additional constraint will reduce the correlation between surface gravity (radius) and temperature of both components.

\section{GSSP}\label{s:gssp}
A spectroscopic analysis using the grid search in stellar parameters (GSSP) package was performed to determine the atmospheric parameters of the cool companions. This analysis was performed on a master spectrum created by shifting all spectra to the rest velocity of the cool companion and then summing them. The lines of the hot component are then smeared out but as they are few, this did not hamper our analyses of the cool component.

\begin{figure*}
    \includegraphics{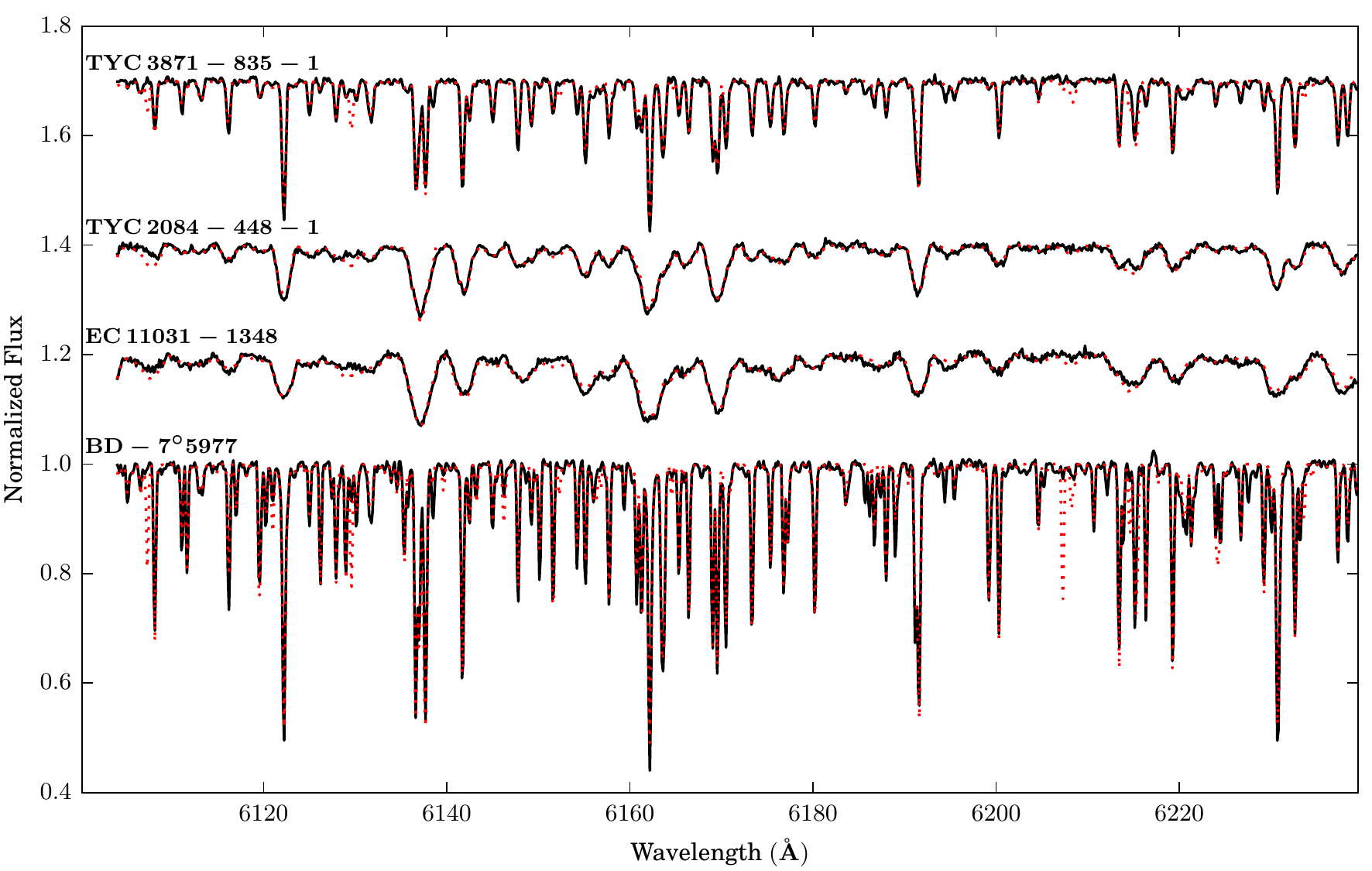}
    \caption{The observed normalised and co-added HERMES spectra (black full line) and the best fitting GSSP model (red dotted line) for a section of the wavelength range used to determine the spectroscopic parameters with the GSSP code.}
    \label{fig:gssp_fit}
\end{figure*}

\subsection{The GSSP code}
The GSSP software package \citep{Tkachenko2015} is based on a grid search in the fundamental atmospheric parameters and (optionally) individual chemical abundances of the star (or binary stellar components) in question. It uses the method of atmosphere models and spectrum synthesis, which assumes a comparison of the observations with each theoretical spectrum from the grid. For the calculation of synthetic spectra, the SynthV LTE-based radiative transfer code \citep{Tsymbal1996} and a grid of atmosphere models pre-computed with the {\sc LL models} code \citep{Shulyak2004} are used.

GSSP allows for optimisation of six stellar parameters at a time: effective temperature (T$_{\rm eff}$), surface gravity ($\log{g}$), metallicity ([Fe/H]), micro-turbulent velocity (v$_{\rm micro}$), projected rotational velocity (\vsini) and the dilution ($F_{\rm MS} / F_{\rm total}$) of the star. The synthetic spectra can be computed in any number of wavelength ranges, and each considered spectral interval can be from a few Angstrom up to a few thousands Angstrom wide. The grid of theoretical spectra is built from all possible combinations of the above mentioned parameters. Each spectrum from the grid is compared to the observed spectrum of the star and the chi-square merit function is used to judge the goodness of fit. The code delivers the set of best fit parameters, the corresponding synthetic spectrum, and the chi-square values for each grid point.

\subsection{Results}
The GSSP analysis was performed on the wavelength range of 5800 - 6500 \AA, as these ranges give the best compromise between a high signal to noise, and a high contribution of the cool companion. The spectral regions in this range that can contain lines from the sdB component are ignored.

The ranges of the model grid used in the analysis are given in Table\,\ref{tb:gssp-grid-ranges}. The micro-turbulent velocity is kept fixed at v$_{\rm micro}$ = 2.0 \kms. The projected rotational velocity is unconstrained and uses steps of 1 km s$^{-1}$. To assure that the global minimum is found, first a coarse grid spanning the entire parameter space is calculated, after which a smaller grid around the best fitting model is calculated using the smallest available step size in each parameter. The errors given are the 1\,$\sigma$ errors, which take into account all possible parameter correlations. The contribution of the sdB component is treated as a wavelength independent dilution factor, which is included in the fit as a free parameter.

A spectral analysis of the cool companion of PG\,1104+243 was already presented in \citet{Vos2012}. The results of the GSSP analysis for BD$-$7$^{\circ}$5977,  EC\,11031--1348, TYC\,2084--448--1 and TYC\,3871--835--1 is given in the bottom half of Table\,\ref{tb:sed_gssp_parameters}. The best fitting GSSP model together with the observed spectra for part of the fitted wavelength range is shown in Fig.\,\ref{fig:gssp_fit}. The results for each system are discussed below.

\begin{description}
 \item[BD$-$7$^{\circ}$5977:] The cool companion is clearly a subgiant with a surface gravity of 3.0 $\pm$ 0.4 dex and a temperature of 4800 $\pm$ 100 K. The subgiant is a slow rotator (\vsini\ = 9 $\pm$ 1 \kms), and the sharp lines help in the determination of the parameters. In the spectral region fitted with GSSP, the subgiant contributes 72\% of the light.\\ 
 
 \item[EC\,11031--1348:] The cool companion is a fast rotator with \vsini\ = 70 $\pm$ 4 \kms. The line blending makes it harder to correctly determine the continuum, which also results in a higher uncertainty on the temperature and dilution factor. The best fitting spectral model corresponds to spectral type F6V at T$_{\rm eff}$ = 6400 $\pm$ 250 K, and it has a $\log{g}$ of 4.3 $\pm$ 0.5 dex. The companion has a slightly super-solar metallicity of [Fe/H] = 0.2 $\pm$ 0.2, and contributes 84\% of the light in the fitted region.\\
 
 \item[TYC\,2084--448--1:] The companion is again a fast rotator with \vsini\ = 52 $\pm$ 3 \kms, and a spectral type of F5V. The effective temperature is 6300 $\pm$ 200 K and it surface gravity is $\log{g} = 4.4 \pm 0.5$~dex. It has a slightly sub-solar metallicity at [Fe/H]~$= -0.1 \pm 0.1$~dex, and contributes 75\% of the light in the fitted range.\\
 
 \item[TYC\,3871--835--1] has a F7V companion with an effective temperature of $6200 \pm 150$~K and a surface gravity of $4.4 \pm 0.3$~dex. The companion has a rotational velocity of $15 \pm 1$~\kms\ and a slightly super-solar metallicity of [Fe/H]~$= 0.2 \pm 0.2$~dex. In the fitted range, it contributes 65\% of the light.\\
\end{description}

As can be seen in Fig.\,\ref{fig:gssp_fit}, there are a few lines visible in the theoretical models which are missing in the observed spectra. This is likely due to errors in the line lists used for the LL models.

For all four systems, the spectral parameters derived with GSSP correspond well with those obtained in the SED fit. However, we do note that the SED fit method consistently results in a higher contribution for the cool companion in the wavelength range 5800--6500 \AA\ than the GSSP fit. As both methods are completely independent, and based on different observations, we can conclude that the spectral parameters derived for the cool companions are reliable.

\begin{table}
   \centering
   \caption{The parameter ranges of the model grid used in the GSSP fit together with the step size. The micro-turbulent velocity is kept fixed.} \label{tb:gssp-grid-ranges}
   \begin{tabular}{lll}
    \hline
    Parameter   &   Range   &    Step \\ \hline \hline
    \noalign{\smallskip}
    T$_{\rm eff}$ (K)            &   3000 -- 9000   &   100  \\
    log $g$ (dex)                &   2.5 -- 4.5     &   0.1  \\
    $[$M/H$]$ (dex)              &   -0.8 -- 0.8    &   0.1  \\
    v$_{\rm micro}$ (\kms)       &   2.0            &   /    \\
    v$_{\rm rot} \sin{i}$ (\kms) &   1 -- 150       &   1    \\
    dilution                     &   0.01 -- 1.00   &   0.01 \\\hline
   \end{tabular}
\end{table}

\section{Wide sdB binary sample}\label{s:wide_sdB_sample}
Including the four new systems presented in this article, there are eleven wide sdB binaries known of which the orbit has been solved. The first such system, PG\,1018$-$047 was presented by \citet{Deca2012}, with updated properties based on UVES spectroscopy published by \citet{Deca2017}. Two systems, PG\,1449+653 and PG\,1701+359, were observed by \citet{Barlow2013}. And four more systems were presented in part 1 and 2 of this series. The main properties of these systems are summarised in Table\,\ref{tb:wide_sdb_overview}.

To derive the inclination of the system and the mass of the cool companion we have assumed a canonical sdB mass of M$_{\rm sdB}\,=\,0.47\,\pm\,0.05 M_{\odot}$. The inclination can then be derived from the $M \cdot \sin{i}$ value obtained from the radial velocity curves as: $ i = \sin^{-1}(M_{\rm sini,sdB} / 0.47)$. The total separation (a) can be derived from the reduced separation using the inclination. The radius of both components is derived using the mass and the surface gravity as $R = \sqrt{G M / g}$. The rotational velocity of the cool companion is corrected for the inclination, by assuming that the rotation axis of the cool companion is perpendicular to the orbital plane. The errors are propagated using a Monte-Carlo simulation.

The distance estimates are calculated from the distance modulus: $\log{d} = (m_V - M_V + 5)/5$. For the systems presented in \citetalias{Vos2012, Vos2013} and this article, the absolute magnitudes ($M_V$) are obtained by using a binary SED model with the best fitting atmospheric parameters determined from the SED and GSSP fit. This model is integrated over the Johnson V band and scaled to a distance of 10 pc. The apparent magnitudes ($m_V$) are taken from literature observations. For PG\,1018-047 the same method is applied \citep{Deca2017}. For PG\,1449+653 and PG\,1701+359 no spectral analysis was performed, and \citet{Barlow2013} estimated the distance from the distance modulus by assuming T$_{\rm eff, sdB} = 24\,000 - 38\,000$ K and R$_{\rm sdB} = 0.15-0.22$ R$_{\odot}$ and taking the conventional correlations between these parameters for sdBs into account. While for the cool companion solar abundance was assumed together with the temperatures and radii corresponding to its spectral type.

\begin{table*}
 \centering
 \caption{The main orbital and spectral properties of all wide sdB binary systems of which the orbits are solved. The parameters of PG\,1018$-$047 are taken from \citet{Deca2017}, those of PG\,1449+653 and PG\,1701+359 are taken from \citet{Barlow2013}, PG\,1104$-$243 from \citet{Vos2012} and this article, BD$+$29$^{\circ}$3070, BD$+$34$^{\circ}$1543 and Feige\,87 are taken from \citet{Vos2013}. The remaining systems given in bold are from this article. The mass of the sdB component is taken at its canonical value of 0.47 M$_{\odot}$. M$_{\rm acc}$ is the mass accreted by the companion during the RLOF phase based on its rotational velocity. $\tau_{\rm acc, min}$ is the minimum time needed to accrete this amount of matter based on the Eddington limit.} \label{tb:wide_sdb_overview}
\begin{tabular}{ll@{ $\pm$ }rl@{ $\pm$ }rl@{ $\pm$ }rl@{ $\pm$ }rl@{ $\pm$ }rl@{ $\pm$ }r}
   \hline \noalign{\smallskip}
   & \multicolumn{2}{c}{PG\,1018--047} & \multicolumn{2}{c}{PG\,1449+653} & \multicolumn{2}{c}{PG\,1701+359} & \multicolumn{2}{c}{PG\,1104+243} & \multicolumn{2}{c}{BD+29$^{\circ}$3070} & \multicolumn{2}{c}{BD+34$^{\circ}$1543} \\\hline\hline
   \noalign{\smallskip}
Period (d)           & 752    & 2      & 909    & 2      & 734    & 3      & 755    & 3      & 1254   & 5      & 972    & 2      \\
Eccentricity         & 0.05   & 0.01   & 0.11   & 0.02   & 0.00   & 0.07   & 0.04   & 0.02   & 0.15   & 0.01   & 0.16   & 0.01   \\
$i$ (dgr)            & 45     & 2      & 63     & 6      &\multicolumn{2}{c}{/}& 32     & 3      & 73     & 3      & 43     & 1      \\
q (M$_{\rm sdB}$/M$_{\rm ms}$)           & 0.7    & 0.02   & 0.64   & 0.06   &\multicolumn{2}{c}{/}& 0.7    & 0.02   & 0.37   & 0.01   & 0.57   & 0.01   \\
a (R$_{\odot}$)      & 367    & 10     & 419    & 19     &\multicolumn{2}{c}{/}& 322    & 12     & 585    & 5      & 447    & 4      \\
d (pc)               & 900    & 100    & 980    & 190    & 690    & 150    & 282    & 15     & 226    & 37     & 207    & 30     \\
\noalign{\smallskip} & \multicolumn{12}{c}{MS component}\\ \noalign{\smallskip}
M (M$_{\odot}$)      & 0.71   & 0.04   & 0.73   & 0.10   &\multicolumn{2}{c}{/}& 0.67   & 0.07   & 1.26   & 0.06   & 0.82   & 0.07   \\
R (R$_{\odot}$)      & 1.2    & 0.3    &\multicolumn{2}{c}{/}&\multicolumn{2}{c}{/}& 0.88   & 0.12   & 1.28   & 0.97   & 1.21   & 0.48   \\
Teff (K)             & 4210   & 280    &\multicolumn{2}{c}{G0V}&\multicolumn{2}{c}{K0V}& 5970   & 200    & 6190   & 420    & 6100   & 300    \\
$\log{g}$ (cgs)      & 4.8    & 0.3    &\multicolumn{2}{c}{/}&\multicolumn{2}{c}{/}& 4.38   & 0.11   & 4.33   & 0.5    & 4.18   & 0.4    \\
$[$Fe/H$]$ (dex)     & -0.85  & 0.25   &\multicolumn{2}{c}{/}&\multicolumn{2}{c}{/}& -0.58  & 0.11   & 0.05   & 0.16   & -0.26  & 0.08   \\
v$_{\rm rot}$ (km s$^{-1}$) & 10     & 2      &\multicolumn{2}{c}{/}&\multicolumn{2}{c}{/}& 10     & 4      & 55     & 5      & 25     & 6      \\
\noalign{\smallskip} & \multicolumn{12}{c}{sdB component}\\ \noalign{\smallskip}
M (M$_{\odot}$)      &\multicolumn{2}{c}{0.47}&\multicolumn{2}{c}{0.47}&\multicolumn{2}{c}{/}&\multicolumn{2}{c}{0.47}&\multicolumn{2}{c}{0.47}&\multicolumn{2}{c}{0.47}\\
R (R$_{\odot}$)      & 0.3    & 0.1    &\multicolumn{2}{c}{/}&\multicolumn{2}{c}{/}& 0.13   & 0.02   & 0.16   & 0.1    & 0.14   & 0.06   \\
Teff (K)             & 29100  & 3000   &\multicolumn{2}{c}{/}&\multicolumn{2}{c}{/}& 33800  & 2500   & 26200  & 3000   & 36600  & 3000   \\
$\log{g}$ (cgs)      & 5.2    & 0.3    &\multicolumn{2}{c}{/}&\multicolumn{2}{c}{/}& 5.94   & 0.2    & 5.69   & 0.45   & 5.84   & 0.35   \\
\noalign{\smallskip} & \multicolumn{12}{c}{RLOF accretion}\\ \noalign{\smallskip}
M$_{\rm acc}$ (M$_{\odot}$) & 0.004  & 0.001  &\multicolumn{2}{c}{/}&\multicolumn{2}{c}{/}& 0.003  & 0.001  & 0.034  & 0.011  & 0.012  & 0.003  \\
$\tau_{\rm acc, min}$ (yr) & 3      & 1      &\multicolumn{2}{c}{/}&\multicolumn{2}{c}{/}& 4      & 1      & 29     & 19     & 10     & 5      \\
\noalign{\smallskip} \hline \noalign{\smallskip}
& \multicolumn{2}{c}{Feige\,87} & \multicolumn{2}{c}{\textbf{BD--7$^{\circ}$5977}} & \multicolumn{2}{c}{\textbf{EC\,11031--1348}} & \multicolumn{2}{c}{\textbf{TYC\,2084--448--1}} & \multicolumn{2}{c}{\textbf{TYC\,3871--835--1}} \\\hline  \noalign{\smallskip}
Period (d)           & 938    & 2      & 1262   & 1      & 1099   & 6      & 1098   & 5      & 1263   & 5      \\
Eccentricity         & 0.11   & 0.01   & 0.16   & 0.01   & 0.17   & 0.03   & 0.05   & 0.03   & 0.16   & 0.02   \\
$i$ (dgr)            & 79     & 3      & 22     & 2      & 57     & 4      & 54     & 3      & 17     & 1      \\
q (M$_{\rm sdB}$/M$_{\rm ms}$)            & 0.55   & 0.01   & 0.4    & 0.1    & 0.36   & 0.02   & 0.51   & 0.02   & 0.54   & 0.02   \\
a (R$_{\odot}$)      & 441    & 3      & 570    & 21     & 540    & 16     & 500    & 14     & 530    & 22     \\
d (pc)               & 376    & 44     & 560    & 80     & 530    & 50     & 345    & 60     & 305    & 50     \\
\noalign{\smallskip} & \multicolumn{10}{c}{MS component}\\ \noalign{\smallskip}
M (M$_{\odot}$)      & 0.84   & 0.07   & 1.18   & 0.32   & 1.31   & 0.16   & 0.92   & 0.10   & 0.87   & 0.10   \\
R (R$_{\odot}$)      & 1.02   & 0.5    & 6.8    & 2.6    & 1.7    & 0.7    & 1.1    & 0.4    & 0.9    & 0.2    \\
Teff (K)             & 5980   & 325    & 4810   & 90     & 6480   & 210    & 6440   & 190    & 6270   & 140    \\
$\log{g}$ (cgs)      & 4.36   & 0.42   & 2.90   & 0.30   & 4.15   & 0.35   & 4.46   & 0.33   & 4.37   & 0.30   \\
$[$Fe/H$]$ (dex)     & -0.5   & 0.18   & -0.2   & 0.1    & 0.2    & 0.2    & -0.1   & 0.1    & 0.1    & 0.1    \\
v$_{\rm rot}$ (km s$^{-1}$) & 8      & 3      & 24     & 3      & 84     & 6      & 63     & 3      & 52     & 5      \\
\noalign{\smallskip} & \multicolumn{10}{c}{sdB component}\\ \noalign{\smallskip}
M (M$_{\odot}$)      &\multicolumn{2}{c}{0.47}&\multicolumn{2}{c}{0.47}&\multicolumn{2}{c}{0.47}&\multicolumn{2}{c}{0.47}&\multicolumn{2}{c}{0.47}\\
R (R$_{\odot}$)      & 0.18   & 0.07   & 0.4    & 0.3    & 0.2    & 0.2    & 0.2    & 0.2    & 0.15   & 0.15   \\
Teff (K)             & 27300  & 2700   & 30000  & 7000   & 28000  & 5000   & 25000  & 7000   & 24000  & 7000   \\
$\log{g}$ (cgs)      & 5.58   & 0.34   & 5.20   & 0.60   & 5.60   & 0.50   & 5.80   & 0.70   & 5.80   & 0.70   \\
\noalign{\smallskip} & \multicolumn{10}{c}{RLOF accretion}\\ \noalign{\smallskip}
M$_{\rm acc}$ (M$_{\odot}$) & 0.004  & 0.001  & 0.030  & 0.008  & 0.067  & 0.021  & 0.027  & 0.008  & 0.021  & 0.006  \\
$\tau_{\rm acc, min}$ (yr) & 4      & 2      & 5      & 3      & 49     & 38     & 29     & 20     & 5      & 3      \\
\hline \noalign{\smallskip}
\end{tabular}

\end{table*}

\subsection{Mass accretion during RLOF}\label{s:rlof_mass_accretion}
Based on the rotational velocity of the companion, one can derive an estimate of how much mass it has accreted during the RLOF phase. This can be done by assuming that the system is synchronised before the onset of mass-loss, and that the currently observed rotational velocity is entirely due to the angular momentum transferred during the RLOF phase. The difference in angular momentum before and after the RLOF phase can be linked to the mass accreted during RLOF based on Keplerian disk mass-transfer. Thus the accreted mass is given by \citet{Vos2017}:
\begin{equation}
  M_{\rm acc} = r_{\rm g} \sqrt{\frac{M_{\rm c} R_{\rm c}^3}{G}} \cdot \left( \frac{v_{\rm rot,f}}{R_{\rm c}} - \frac{2 \pi}{P_{\rm orb,i}} \right).\label{eq:accreted_mass}
\end{equation}
Where $r_{\rm g} = 0.076$ is the gyration radius based on polytropic models \citep{Claret1989}, $M_{\rm c}$ and $R_{\rm c}$ are respectively the mass and radius of the cool companion, $G$ is the gravitational constant, $v_{\rm rot,f}$ is the observed rotational velocity of the companion and $P_{\rm orb,i}$ is the estimated initial orbital period. As the binary would have been circularised and synchronised before the onset of RLOF, the initial spin period equals the orbital period before RLOF $P_{\rm spin, i} = P_{\rm orb, i}$. The orbital period before RLOF is estimated between 500 and 900 days, as this corresponds to the separation necessary to initiate RLOF near the tip of the red giant branch \citep{Vos2015}

After the RLOF phase the synchronisation is negligible, as shown by \citet{Vos2017}. Other potential sources of angular momentum loss, such as stellar winds or magnetic fields, are ignored
in the current treatment, as these are difficult to constrain and it is not clear if they are active at all.

A lower limit for the duration of RLOF can be given by calculating the maximum rate of mass accretion for the cool companion based on the Eddington luminosity. This Eddington accretion limit is \citep{Vos2017}:
\begin{equation}
 \dot{M}_{\rm edd} = \frac{4 \pi c}{\kappa} R_{\rm c},
\end{equation}
where c is the speed of light and $\kappa$ is the opacity which in the simplest case can be taken as Thompson scattering $\kappa$ = 0.4 cm$^2$ g$^{-1}$.

The resulting estimates of the accreted mass and the duration of the RLOF phase are given in Table\,\ref{tb:wide_sdb_overview}. For all of the systems for which these estimates could be calculated, the total amount of accreted mass necessary to reach the observed rotational velocity is small, on the order of $10^{-3} - 10^{-2}$ M$_{\odot}$. The calculated timescales based on the Eddington luminosity are on the order of a few years to a few decades. This is an lower limit and the real duration can be much longer, but these timescales do allow enough mass to be accreted onto the cool companions from fast thermal-timescale mass-loss of a red giant during RLOF. For example, models calculated with the MESA \citep{Paxton2011, Paxton2013, Paxton2015} stellar/binary evolution code predict a duration for the mass-loss phase of a few hundred years in the case of stable RLOF from a red giant \citep{Vos2015}.

\section{The period-eccentricity distribution}\label{s:period_ecc_distribution}

\begin{figure*}
    \includegraphics{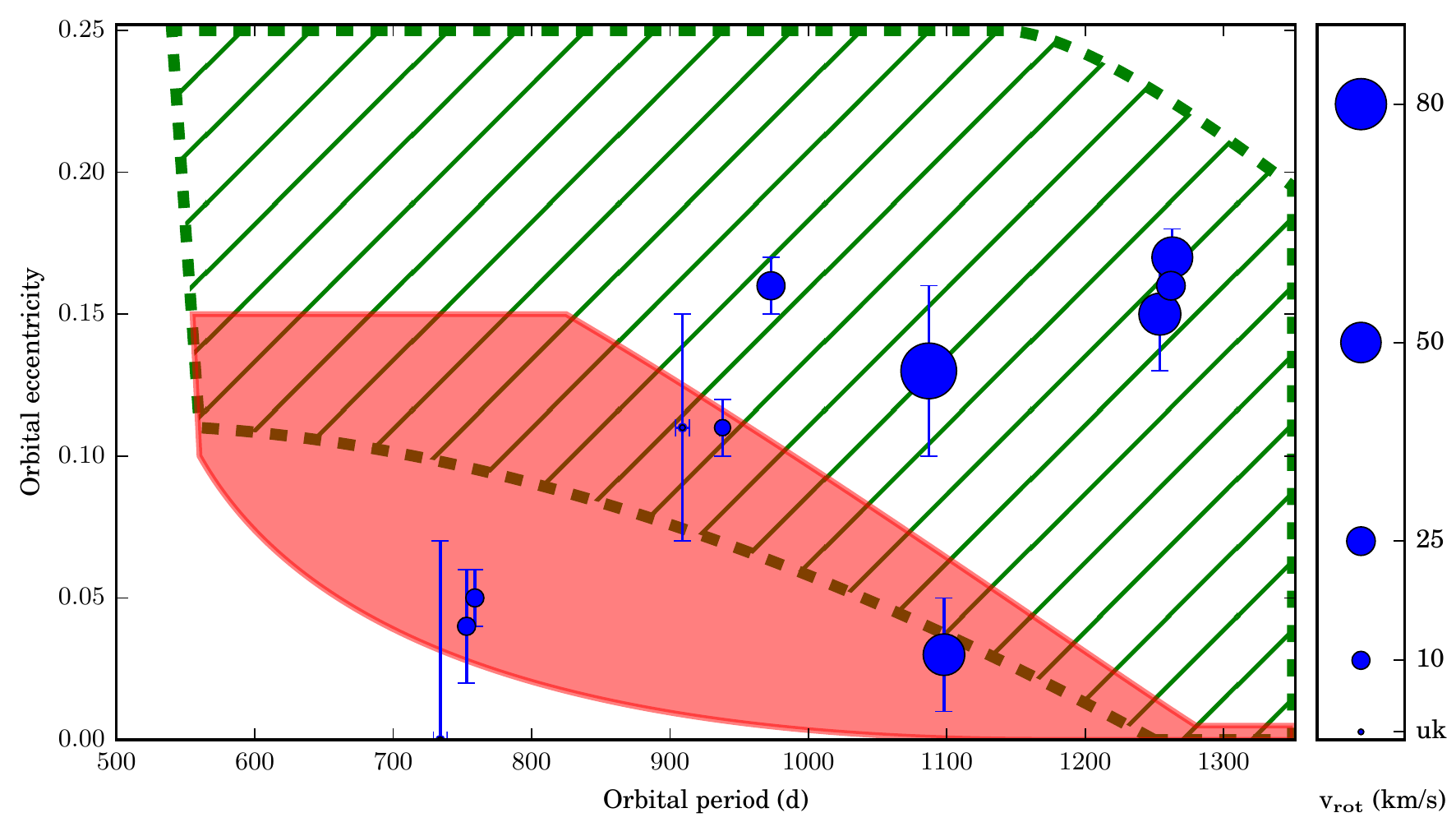}
    \caption{The period and eccentricity of all wide sdB binaries with known orbital parameters is plotted in blue circles. The size of the circle indicates the rotational velocity of the cool companion star. Two stars for which no rotational velocity is known are plotted with blue dots ($v_{\rm rot}$ = uk). The red and green areas indicate the period--eccentricity range that can be covered by the MESA binary evolution models with eccentricity pumping included as presented in \citet{Vos2015}. Models only including phase-dependent mass-loss are shown in red shaded full border, while models that also include a CB disk are shown in green hatched dotted border. The range of the MESA models shown here is truncated at a maximum orbital period of 1350 days and a maximum eccentricity of 0.25.  }
    \label{fig:pe_distribution}
\end{figure*}

The period-eccentricity distribution of the eleven systems given in Table\,\ref{tb:wide_sdb_overview} is shown in Fig.\,\ref{fig:pe_distribution}. As can be seen on this figure, the majority of the systems has a significant non-zero eccentricity. Only one of the eleven systems, PG\,1701+359, has a circular orbit. \citet{Barlow2013} indicate that the orbit of PG\,1701+359 can be fitted with an eccentricity of 0.07, if the eccentricity is left as a free parameter. The \citet{Lucy1971} test however, indicates that the eccentricity of the orbit is not significant with respect to the errors on the radial velocities. \citet{Barlow2013} furthermore indicate that there are not enough radial velocity measurements on the up-going slope of the radial velocity curve to completely rule out an eccentric orbit, making further observations needed to discern between an eccentric or circular orbit. Two other systems on the shorter end of the orbital period range, PG\,1104+243 \citep{Vos2012} and PG\,1018--047 \citep{Deca2012}, were originally found to have circular orbits, but continued observing with HERMES (PG\,1104+243, this article) and higher S/N observations with UVES (PG\,1018--047, \citealt{Deca2017}) has shown that the orbits of both binaries have low albeit significantly non-zero eccentricities.

A first interesting observation is that none of the systems have orbital periods shorter than 700 days. This is not an observational bias, as systems with shorter orbital period are easier to detect spectroscopically. This supports that there is a clear distinction between the sdB binaries that form through the common envelope ejection channel and end up on a short orbital period (P $\leqslant$ 50 days, \citealt{Kupfer2015}), and the wide sdB binaries that are formed through the stable RLOF channel. The observed lower limit on the orbital period for post stable-RLOF sdB binaries can be used to test stability criteria for RLOF. Contrary to the lower limit, no upper limit can be placed on the orbital periods as the currently observed longest periods are on the same order as the duration of the observing programs, and longer period systems would be harder to detect as the radial velocity amplitudes are lower.

Secondly, there is a trend of higher eccentricity at higher orbital period, with the short period (around 700 days) systems having eccentricities between 0 and 0.05, while systems with longer orbital periods have eccentricities that reach 0.17. The Pearson test \citep{Pearson1896} was used to calculate the correlation between these two variables. It is defined as the covariance of the two variables divided by their standard deviation. In the case of the period and eccentricity, the Pearson test yields a positive correlation of $r(P,e) = 0.75$ with a 99.2 \% confidence level ($p$ value of 0.008), indicating that the correlation is significant. The only exception to this trend is TYC\,2048--448--1, which has an orbital period of 1098 $\pm$ 5 days and an eccentricity of 0.05 $\pm$ 0.03, while other systems with similar orbital periods have eccentricities above 0.10.

An alternative way of interpreting the period--eccentricity diagram is not as a linear relation, but as two separate populations. Possibly one population at shorter orbital period, and another at longer orbital periods and higher eccentricities. These two population can both be formed by the stable RLOF channel, but there could be other eccentricity pumping mechanisms at play. With the current observed sample, it is not possible to distinguish between both options.

Considering all systems, the highest eccentricity reached by any system is around 0.16-0.17. This might indicate that there is an upper limit on the eccentricity that can be obtained by the eccentricity pumping mechanisms at play during the formation of these wide sdB binaries.

\subsection{Theoretical predictions for wide sdBs}
The eccentricity of these orbits is unexpected as the sdBs are formed through stable RLOF from a red giant. Current circularisation theory predicts all of these systems to be circularised well before the onset of mass loss. Several eccentricity pumping mechanisms are described in the literature. Two of those mechanisms, phase-dependent mass-loss during RLOF \citep[e.g.][]{Eggleton2006, Bonacic2008} and the formation of a circumbinary (CB) disk during RLOF \citep[e.g.][]{Artymowicz1994}, are capable of explaining the current observed population \citep{Vos2015}. Phase-dependent mass-loss during RLOF can increase the eccentricity as more mass will be lost near periastron than near abastron. A CB disk can form during the RLOF phase if not all the mass is accreted by the companion. This is likely the case, as the sdB progenitor needs to lose at least half a solar mass during the RLOF phase on a short timescale. A CB disk can increase the eccentricity of the orbit, as well as slightly reduce the orbital period due to Lindbladt resonances. 

Currently there is no direct observational evidence for the existance of a CB disk around sdB binaries. The presence of such a disc is inspired by the eccentric post-AGB binaries where such discs are commonly observed \citep[e.g.][see also Sect.\,\ref{s:pe_postAGB}]{VanWinckel2003, Gezer2015}. Further support for CB disks around sdB binaries is that the sdB progenitor has to lose at least 0.5\,$M_{\odot}$ during the RLOF phase, most of which is not accreted by the companion.

The period--eccentricity coverage of binary models including these eccentricity pumping mechanisms calculated by \citet{Vos2015} is shown in Fig.\,\ref{fig:pe_distribution}. There are two sets of models, the first with only eccentricity pumping due to phase dependent RLOF (shown in red shaded area), and the second set with both phase dependent RLOF and a CB disk (shown in green hatched area). These models do clearly allow for all of the observed systems, with the exception of the circular short period orbit of PG\,1701+359. However, these models are not capable of explaining the period--eccentricity trend. Contrary to the observations, the eccentricity pumping mechanisms are more efficient at shorter orbital periods, and still predict circularised orbits at longer orbital periods.

A combination of both models with only phase-dependent RLOF and models that also include a CB disk is necessary to explain all observed wide sdB binaries. This would support the two population view of the observed period--eccentricity distribution. However, it has to be noted that the goal of \citet{Vos2015} was to find models that would allow for the observed systems, and not to find the full extent of the parameter space that can be covered by both models.

These eccentricity pumping mechanisms and the principles of mass-loss during stable-RLOF in general are not well understood, and are parameterised in most evolution codes. For example, the way mass is lost from the system during stable RLOF in MESA is determined by the formalism of \citet{Tauris2006} who defined three different mass-loss fractions. These fractions are treated as free parameters in the MESA models, and have a strong influence on the final orbital period. Similarly, physical properties of CB disks such as the total mass that they can contain, or their lifetime are not known from observations, and are treated as free parameters.

To link the parameterised interaction mechanisms to system properties, and create theoretical binary evolution models that are capable to predict the period--eccentricity distribution of wide sdB binaries, it is important to understand how the eccentricity and orbital period depend on other properties of the systems. 

\subsection{Period-eccentricity of post-AGB systems}\label{s:pe_postAGB}
\begin{figure}
    \includegraphics{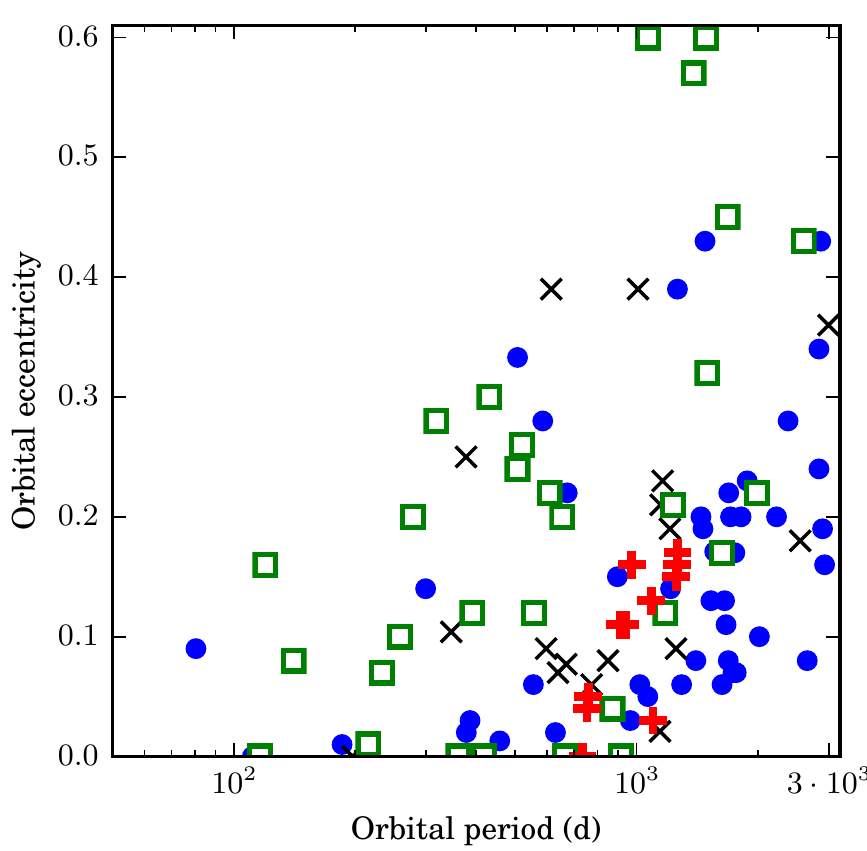}
    \caption{The period--eccentricity distribution of different types of post-AGB binaries that have undergone a mass-loss phase in their past: dusty pAGB binaries (green squares, \citealt{Waters1993, VanWinckel1995, Pollard1996, Gonzalez1996, VanWinckel1998, VanWinckel1999, VanWinckel2000, Maas2002, Maas2005, VanWinckel2007, VanWinckel2009, Gorlova2014, Gezer2015, DeSmedt2016, Manick2017}), Ba stars (blue circles, \citealt{Vanderswaelmen2017}), Tc-poor S stars (black x's, \citealt{Vanderswaelmen2017}) and the wide sdB binaries (red crosses, this article). See section \ref{s:pe_postAGB}.}
    \label{fig:pe_postAGB}
\end{figure}

In Fig.\,\ref{fig:pe_postAGB} the period--eccentricity distribution of post-AGB (pAGB) binaries are plotted. This sample contains dusty pAGB binaries with known orbital parameters from \citet{Waters1993, VanWinckel1995, Pollard1996, Gonzalez1996, VanWinckel1998, VanWinckel1999, VanWinckel2000, Maas2002, Maas2005, VanWinckel2007, VanWinckel2009, Gorlova2014, Gezer2015, DeSmedt2016, Manick2017}, and  barium (Ba) and Tc-poor S stars for which orbital parameters were derived by \citet{Vanderswaelmen2017}. The Ba stars are giant stars in binary systems with a WD companion where the giant has an overabundance in s-process elements caused by mass transfer from WD progenitor during its AGB phase \citep{Bidelman1951, Jorissen1998}. The Tc-poor S stars in this sample are extrinsic S-star part of binary systems for which the over abundance in s-process elements is caused by mass transfer in a similar fashion as the Ba-stars \citep{Jorissen1998}.

The pAGB systems reach orbital periods in excess of 10\,000 days.  Here we only discuss the pAGB binaries with orbital period shorter than 3000 days as those systems should have circularised due to tides \citep{Pols2003}. Eccentric pAGB systems with orbital periods longer then 3000 days could have kept their initial eccentricity, and are for that reason not interesting in this comparison.

The range in both orbital period and eccentricity for the pAGB systems is larger than for the wide sdB binaries. The observed periods vary from 100 days to several 1000 days, and the eccentricities can go up to 0.6. Compared to a period range of 700 - 1300 days and a maximum eccentricity of 0.17 for sdB binaries. The fact that for wide sdB binaries, the mass of the sdB star has a lower limit at the He-ignition core mass, and at the same time needs to have lost almost the entire H envelope, limits the period range in which these systems can form. These strict limits do not exist in the pAGB binaries, making the sdB wide binaries a perfect sample to test binary interaction mechanisms.

Both the period-eccentricity distribution of the wide sdBs and the post-AGB systems show a trend of higher possible eccentricities at higher orbital periods, indicating an upper limit on the eccentricity depending on the orbital period. The Pearson correlations for the different types of post-AGB binaries between period and eccentricity are $r(P,e) = 0.46$ with a $p$ value of 0.001 for the Ba systems, $r(P,e) = 0.41$ with a $p$ value of 0.07 for the S-type binaries and $r(P,e) = 0.57$ with a $p$ value of 0.001 for the dusty pAGB binaries.

The spread in eccentricity for the pAGB systems is larger than for the wide sdB binaries, but there is a striking similarity in the lacking of circular systems at longer orbital periods. For the pAGB systems, circularised systems exist up to $P \sim 1000$ days. At longer orbital periods the minimum eccentricity increases to $e \sim 0.1$. This can be compared to the wide sdB binaries where there is one system with a longer orbital period and a lower eccentricity, TYC\,2048--448--1, while the other systems with $P \gtrsim 1000$ d all have eccentricities larger then 0.1.

The pAGB systems are capable of reaching higher eccentricities than the wide sdB binaries. If only systems with P < 3000 days are considered, 43 \% of the post-AGB binaries have a higher eccentricity than the most eccentric wide sdB-binary. This indicates that the eccentricity pumping mechanisms in the post AGB binaries are more efficient or are effective over a longer timescale.

The possibility of having two separate populations of wide sdB binaries that was discussed in the previous section is not immediately visible for the pAGB binaries. This does not prevent the existence of two populations formed by, for example, different eccentricity pumping mechanisms. The continuous distribution in both period and eccentricity would, however, indicate that the period - eccentricity ranges reached by each population overlap.

\subsection{Rotational velocity - accreted mass}
A first observation is that the rotational velocity of the cool companions increases for the longer period systems. The two short period systems for which the rotational velocity is determined, have $v_{\rm rot} \sim$ 10 km s$^{-1}$, while the systems with a longer orbital period have rotational velocities reaching 80 km s$^{-1}$ or more. The SG component of BD$-$7$^{\circ}$5977 has a rotational velocity of 24 $\pm$ 3 km s$^{-1}$, which is lower then the two other systems with orbital period around 1250 days. But the SG companion will have a decreased rotational velocity due to its inflated atmosphere compared to the MS companions of the other two systems. 

The rotational velocity of the cool companion can be linked to the amount of mass accreted during the RLOF phase as calculated in Section\,\ref{s:rlof_mass_accretion}. The relation of a higher rotational velocity of the companion in systems with a longer orbital period indicates that those systems have accreted more mass during the mass-loss phase of the sdB progenitor. The Pearson correlation between the period and rotational velocity of the companion calculated for the 9 systems for which $v_{\rm rot}$ is known is $r(P, v_{\rm rot}) = 0.57$ with a $p$ value of 0.10, indicating a moderate correlation. 

In the view that there is not a continuous relation between the orbital period and the eccentricity / rotational velocity, but that there are separate populations, the division between the systems containing a companion with a higher rotational velocity and the systems containing a companion with a slower rotational velocity roughly corresponds with the division between the two types of eccentricity pumping MESA models. Systems which have less matter accreted onto the companion fall in the period-eccentricity range covered by models with only phase-dependent mass-loss. The systems with a faster rotating companion correspond with the models that have both phase-dependent mass-loss and have had a CB disk. The exception again being TYC\,2084--448--1, which has a fast rotating companion but has an eccentricity that falls just below the range covered by the models including a CB disk.

The MESA models of \citet{Vos2015} were calculated without tracking the rotational velocity of the components. Future models would benefit from including this in the evolution.

\subsection{Initial mass-ratio}
A system property that is slightly harder to determine is the initial mass-ratio before the onset of mass-loss. For most systems this can not be determined without having accurate models of the evolutionary history of the system, which are of course not available in this case. However, there are two type of systems for which it is possible to make an estimate on the initial mass ratio. The first being BD$-$7$^{\circ}$5977, which has a SG companion, meaning it has evolved off the main sequence. Because the lifetime of the sdB star on the horizontal branch is short (estimated at $\sim$ 100 Myr) this evolution would only be possible if the subgiant would be very close in mass to the sdB progenitor. This is only possible if the initial mass-ratio for this system $q_i = M_{\rm MS} / M_{\rm p-sdB}$ is very close to 1.

The second type of systems are those with an M or K type companion. Even though the exact initial mass-ratio of these systems can not be determined, it will be significantly lower than unity. In the current sample there are two sdB binaries with a dwarf companion, PG\,1018--047 and PG\,1701+359, which both have K-type companions. The systems with an SG companion and those with a M/K type companion respectively represent the upper and lower range in initial mass-ratio.

The current sample of wide sdB binaries only contains three system falling in these groups. We note here that the sdB+K binaries with a low initial mass ratio, are also the systems with the shortest orbital periods and lowest eccentricities. Contrary, the sdB+SG binary with $q_i \approx 1$, is one of the systems with the longest orbital periods and higher eccentricities. With only 3 systems we can’t draw any firm conclusions, but they do give us a range of possible initial mass-ratios. Discovering new sdB systems with dwarf or sub-giant companions would be highly valuable as it will allow to study the effect of the initial mass-ratio on the final period-eccentricity.

\section{Summary and Conclusions}
Radial velocities were derived from high resolution spectra taken with the HERMES spectrograph as part of a long term monitoring program. The radial velocity curves obtained allowed us to determine accurate orbital parameters of 4 new systes: BD$-$7$^{\circ}$5977, EC\,11031--1348, TYC\,2084--448--1 and TYC\,3871--835--1. All of these systems have orbital period in excess of 1000 days.  BD$-$7$^{\circ}$5977, EC\,11031--1348 and TYC\,3871--835--1 have a higher eccentricity of $\sim 0.16$ while TYC\,2084--448--1 has an eccentricity of 0.05 $\pm$ 0.03. In addition, the orbital parameters of PG\,1104+243 have been updated based on our new spectra, and this system has now a small but non-zero eccentricity of 0.04 $\pm$ 0.02. 

Literature photometry was used to construct photometric SEDs for the four new systems, which were fitted with atmosphere models to derive the effective temperatures and surface gravities of both the sdB and the cool companion. Furthermore the HERMES spectra were analysed with the GSSP code to obtain an independent second set of atmospheric parameters for the cool companions. BD$-$7$^{\circ}$5977 has a subgiant companion with a surface gravity of 2.9 $\pm$ 0.30, while the other three cool companions are all early F-type stars. The effective temperature of the hot subdwarfs vary between 24\,000 and 30\,000 K, but it has to be noted that the errors on these temperatures are large, around 3000 K. When GAIA parallaxes become available for these systems, the distance to the system can be incorporated as an extra constraint which will greatly reduce the uncertainty on the parameters derived from the photometric SEDs.

The entire wide-sdB binary population with solved orbits consists of eleven systems. Eight of these systems are part of the long term observing campaign with HERMES and have been presented in \citetalias{Vos2012, Vos2013} and this article. The first wide sdB system to be discovered was PG\,1018-047 \citep{Deca2012}, and two more systems, PG\,1449+653 and PG\,1701+359, were presented by \citet{Barlow2013}. Of the latter two systems only an orbital solution and an estimate of the companion type is available, while for the other 9 systems the spectral parameters of both the sdB and the cool companion are known.

The period-eccentricity distribution of the complete sample of wide sdB binaries shows a trend of higher eccentricities at longer orbital periods, varying from eccentricities < 0.05 at periods around 750 days to $e \sim 0.16$ at $P \sim 1250$ days. There is however one exception to this trend, TYC\,2084--448--1, which has an eccentricity of 0.05 $\pm$ 0.03 at $P \sim 1098 \pm 5$ days. Furthermore, there seems to be a lower limit on the orbital period of wide sdB binaries around 700 days. This lower limit on the period can be used to determine the stability criteria of RLOF.

The period-eccentricity distribution observed for the post-RGB wide sdB binaries has several similarities to the distribution observed for post-AGB Ba-star and TC-poor S-type binaries. Even though the range of both period and eccentricity for the pAGB systems is larger then for the wide sdBs, it shows the same trend of higher eccentricities at longer orbital periods. Furthermore, the pAGB systems show a lower limit on the orbital period at $P \approx 1000$ days, over which no circularised systems exist anymore. The wide sdB population shows a similar lack of circularised systems at longer orbital periods.

These wide sdB binaries are post stable-RLOF systems, and by assuming a synchronised state before the onset of mass-loss, the rotational velocity of the companion together with its radius and orbital properties can be used to estimate the amount of mass that was accreted during the RLOF phase. The accreted mass estimate is only a fraction of the mass that the sdB progenitor has to lose, on the order of $10^{-3} - 10^{-2}$ solar masses. When using the Eddington luminosity to derive the minimum time necessary to accrete this mass, this time varies from a few years to a few decades. The amount of mass accretion corresponds to the evolved state of some of the companions, which would not be possible to attain if they have accreted a large amount of mass during RLOF \citepalias{Vos2012}, and the minimum timescales allow for the short periods of mass loss predicted by the MESA models of \citet{Vos2015}.

When comparing theoretical models with the observed population it is clear that models with eccentricity pumping mechanisms activated allow for the existence of most of the observed systems. The only exception being the short period circular system PG\,1701+359. However, the models are not capable of predicting the observed trend of higher eccentricities at higher orbital periods. By comparing the spectral properties with the orbital parameters, we find that there is a moderate correlation of rotational velocity, and thus accreted mass, with orbital period. Furthermore, it is possible that the initial mass ratio could be related to the final period and eccentricity. The inclusion of such possible property correlations can improve the theoretical models, and help explain the observed trends. However, the current sample of wide sdB binaries is too small to allow for the derivation of statistically valid correlations.

\begin{acknowledgements}
Based on observations made with the Mercator Telescope, operated on the island of La Palma by the Flemmish Community, at the Spanish Observatorio del Roque de los Muchachos of the Instituto de Astrof\'{\i}sica de Canarias. 
Based on observations obtained with the HERMES spectrograph, which is supported by the Research Foundation - Flanders (FWO), Belgium, the Research Council of KU Leuven, Belgium, the Fonds National de la Recherche Scientifique (F.R.S.-FNRS), Belgium, the Royal Observatory of Belgium, the Observatoire de Gen\`{e}ve, Switzerland and the Th\"{u}ringer Landessternwarte Tautenburg, Germany. 
JV acknowledges financial support from FONDECYT in the form of grant number 3160504.
HVW and JV acknowledge support from the Research Council of the KU Leuven under grant number GOA/2013/012
This publication makes use of data products from the Two Micron All Sky Survey, which is a joint project of the University of Massachusetts and the Infrared Processing and Analysis Centre/California Institute of Technology, funded by the National Aeronautics and Space Administration and the National Science Foundation.
This research was made possible through the use of the AAVSO Photometric All-Sky Survey (APASS), funded by the Robert Martin Ayers Sciences Fund.
The authors thank all contributing observers to the long-term radial velocity monitoring programme.
\end{acknowledgements}

\bibliographystyle{aa}
\bibliography{bibliography}

\begin{appendix}
\section{Radial velocity tables}\label{a:rv_tables}

\begin{table}
\caption{The radial velocities of both components of BD$-$7$^{\circ}$5977}
\label{tb:BD-7_rv}
\centering
\begin{tabular}{lrrlrr}
\hline\hline
\noalign{\smallskip}
   \multicolumn{3}{c}{MS component}              &   \multicolumn{3}{c}{sdB component}             \\
   BJD       &   RV            &   Error         &   BJD       &   RV            &   Error         \\
   -2450000  &   km s$^{-1}$   &   km s$^{-1}$   &   -2450000  &   km s$^{-1}$   &   km s$^{-1}$   \\\hline
\noalign{\smallskip}
5039.62  &   -5.757   &   0.002  &  5053.78  &  -10.6   &   1.8  \\
5051.66  &   -5.732   &   0.003  &  5095.55  &  -12.4   &   1.3  \\
5052.71  &   -5.780   &   0.002  &  5431.04  &   -4.4   &   1.3  \\
5052.71  &   -5.780   &   0.002  &  5480.84  &   -4.2   &   1.2  \\
5052.71  &   -5.780   &   0.002  &  5779.19  &   -0.6   &   1.1  \\
5063.58  &   -5.665   &   0.002  &  5849.24  &    2.8   &   1.0  \\
5063.58  &   -5.665   &   0.002  &  5938.81  &   -1.4   &   2.5  \\
5090.59  &   -5.682   &   0.004  &  6126.33  &    0.4   &   2.0  \\
5090.59  &   -5.682   &   0.004  &  6173.63  &   -9.5   &   2.4  \\
5100.53  &   -5.802   &   0.002  &  6265.43  &   -8.3   &   6.1  \\
5100.53  &   -5.802   &   0.002  &  6461.70  &   -9.4   &   1.6  \\
5398.71  &   -8.295   &   0.005  &  6535.60  &  -12.5   &   3.4  \\
5434.62  &   -8.724   &   0.003  &  6613.07  &   -7.1   &   1.4  \\
5440.64  &   -8.768   &   0.001  &  6832.36  &   -3.8   &   0.6  \\
5440.64  &   -8.768   &   0.001  &  6899.61  &    1.4   &   1.9  \\
5440.64  &   -8.768   &   0.001  &  6974.42  &   -3.2   &   1.2  \\
5471.52  &   -9.119   &   0.003  &  7195.70  &    8.4   &   2.6  \\
5471.52  &   -9.119   &   0.003  &  7277.59  &    1.7   &   2.6  \\
5499.49  &   -9.376   &   0.003  &                               \\
5767.69  &   -10.856  &   0.002  &                               \\
5790.69  &   -10.917  &   0.002  &                               \\
5831.54  &   -11.016  &   0.003  &                               \\
5843.46  &   -11.029  &   0.002  &                               \\
5859.52  &   -10.847  &   0.003  &                               \\
5862.44  &   -10.952  &   0.003  &                               \\
5935.31  &   -10.620  &   0.002  &                               \\
5942.31  &   -10.561  &   0.004  &                               \\
6100.69  &   -8.653   &   0.002  &                               \\
6137.65  &   -7.953   &   0.001  &                               \\
6140.66  &   -8.008   &   0.001  &                               \\
6173.63  &   -7.434   &   0.001  &                               \\
6265.43  &   -6.196   &   0.002  &                               \\
6461.70  &   -6.213   &   0.002  &                               \\
6535.60  &   -6.976   &   0.002  &                               \\
6611.43  &   -7.874   &   0.005  &                               \\
6612.41  &   -7.968   &   0.005  &                               \\
6615.38  &   -7.895   &   0.004  &                               \\
6831.69  &   -9.924   &   0.003  &                               \\
6831.69  &   -9.924   &   0.003  &                               \\
6833.70  &   -9.919   &   0.004  &                               \\
6899.61  &   -10.352  &   0.002  &                               \\
6954.51  &   -10.640  &   0.004  &                               \\
6994.33  &   -10.785  &   0.003  &                               \\
7195.70  &   -10.468  &   0.003  &                               \\
7277.59  &   -9.791   &   0.002  &                               \\
\hline
\end{tabular}
\end{table}


\begin{table}
\caption{The radial velocities of both components of EC\,11031--1348}
\label{tb:EC11031_rv}
\centering
\begin{tabular}{lrrlrr}
\hline\hline
\noalign{\smallskip}
   \multicolumn{3}{c}{MS component}              &   \multicolumn{3}{c}{sdB component}             \\
   BJD       &   RV            &   Error         &   BJD       &   RV            &   Error         \\
   -2450000  &   km s$^{-1}$   &   km s$^{-1}$   &   -2450000  &   km s$^{-1}$   &   km s$^{-1}$   \\\hline
\noalign{\smallskip}
5616.60  &  -11.718  &   0.239  &  5644.23  &  -10.61  &  0.69  \\
5648.52  &  -13.404  &   0.230  &  5666.41  &   -6.09  &  1.53  \\
5661.45  &  -14.055  &   0.129  &  5952.00  &    0.36  &  0.77  \\
5938.72  &  -16.535  &   0.219  &  6017.03  &    2.09  &  1.11  \\
5951.65  &  -16.380  &   0.184  &  6069.39  &   -4.05  &  1.08  \\
5980.61  &  -18.483  &   0.211  &  6324.17  &  -19.76  &  0.62  \\
6006.56  &  -18.342  &   0.221  &  6403.46  &  -26.52  &  0.93  \\
6027.51  &  -17.795  &   0.310  &  6613.73  &  -16.17  &  2.02  \\
6069.39  &  -17.376  &   0.248  &  6685.67  &  -13.23  &  1.04  \\
6297.72  &  -10.251  &   0.325  &  6743.07  &   -3.05  &  1.57  \\
6316.36  &   -9.408  &   0.131  &  6800.42  &   -5.02  &  2.70  \\
6333.66  &  -10.225  &   0.236  &  7026.73  &   -0.79  &  2.02  \\
6345.58  &   -9.328  &   0.253  &  7112.21  &    1.15  &  1.18  \\
6389.50  &   -7.841  &   0.291  &  7394.73  &  -16.25  &  1.67  \\
6417.42  &   -8.345  &   0.183  &  7476.52  &  -22.10  &  2.06  \\
6613.73  &   -9.549  &   0.276  &                               \\
6663.69  &   -9.732  &   0.278  &                               \\
6707.65  &  -12.172  &   0.294  &                               \\
6721.65  &  -12.350  &   0.433  &                               \\
6764.50  &  -15.565  &   0.298  &                               \\
6809.39  &  -16.621  &   0.286  &                               \\
7008.76  &  -18.120  &   0.325  &                               \\
7035.71  &  -17.158  &   0.510  &                               \\
7082.62  &  -17.854  &   0.442  &                               \\
7114.81  &  -14.718  &   0.246  &                               \\
7131.41  &  -18.097  &   0.314  &                               \\
7394.73  &  -12.755  &   0.348  &                               \\
7465.56  &   -8.586  &   0.378  &                               \\
7487.49  &  -10.001  &   0.348  &                               \\
\hline
\end{tabular}
\end{table}


\begin{table}
\caption{The radial velocities of both components of TYC\,2084--448--1}
\label{tb:TYC2084_rv}
\centering
\begin{tabular}{lrrlrr}
\hline\hline
\noalign{\smallskip}
   \multicolumn{3}{c}{MS component}              &   \multicolumn{3}{c}{sdB component}             \\
   BJD       &   RV            &   Error         &   BJD       &   RV            &   Error         \\
   -2450000  &   km s$^{-1}$   &   km s$^{-1}$   &   -2450000  &   km s$^{-1}$   &   km s$^{-1}$   \\\hline
\noalign{\smallskip}
6320.76  &  -18.107  &  0.123  &   5641.33  &  -24.56   &   0.68   \\
6836.67  &  -11.463  &  0.040  &   5722.57  &  -23.17   &   0.90   \\
6764.13  &   -9.367  &  0.083  &   5777.99  &  -19.72   &   0.76   \\
6427.13  &  -13.890  &  0.057  &   5828.37  &  -15.35   &   0.81   \\
7196.55  &  -22.052  &  0.094  &   5983.23  &   -6.33   &   0.72   \\
7107.70  &  -20.864  &  0.140  &   6061.20  &   -2.84   &   0.72   \\
6804.04  &  -10.820  &  0.056  &   6117.13  &   -0.52   &   0.78   \\
7227.43  &  -21.826  &  0.094  &   6190.40  &   -2.71   &   0.84   \\
6132.62  &  -22.651  &  0.103  &   6320.76  &   -7.26   &   0.66   \\
6536.37  &  -11.872  &  0.115  &   6396.14  &  -13.02   &   0.73   \\
5653.64  &  -10.056  &  0.032  &   6443.24  &  -14.56   &   0.41   \\
7083.76  &  -20.537  &  0.141  &   6536.37  &  -21.59   &   1.25   \\
7444.70  &  -18.112  &  0.104  &   6695.73  &  -30.16   &   1.00   \\
5701.45  &  -10.500  &  0.115  &   6775.30  &  -24.03   &   1.01   \\
6050.52  &  -21.024  &  0.150  &   6832.30  &  -23.29   &   0.51   \\
5753.14  &  -11.234  &  0.040  &   7095.73  &   -4.05   &   0.80   \\
6007.69  &  -20.624  &  0.102  &   7171.86  &   -1.00   &   0.45   \\
7159.52  &  -21.636  &  0.082  &   7241.92  &   -0.27   &   0.88   \\
5616.70  &   -9.488  &  0.057  &   7464.68  &  -10.35   &   0.59   \\
6092.11  &  -21.924  &  0.085  &   7502.56  &  -13.99   &   1.17   \\
7484.66  &  -17.078  &  0.414  &                                   \\
5793.39  &  -12.866  &  0.098  &                                   \\
6374.70  &  -16.339  &  0.122  &                                   \\
6444.56  &  -14.164  &  0.035  &                                   \\
5958.77  &  -18.679  &  0.067  &                                   \\
6695.73  &   -9.303  &  0.163  &                                   \\
5828.37  &  -14.516  &  0.052  &                                   \\
7502.56  &  -14.813  &  0.137  &                                   \\
7256.42  &  -22.575  &  0.127  &                                   \\
6190.40  &  -21.735  &  0.054  &                                   \\
\hline
\end{tabular}
\end{table}


\begin{table}
\caption{The radial velocities of both components of TYC\,3871--835--1}
\label{tb:TYC3871_rv}
\centering
\begin{tabular}{lrrlrr}
\hline\hline
\noalign{\smallskip}
   \multicolumn{3}{c}{MS component}              &   \multicolumn{3}{c}{sdB component}             \\
   BJD       &   RV            &   Error         &   BJD       &   RV            &   Error         \\
   -2450000  &   km s$^{-1}$   &   km s$^{-1}$   &   -2450000  &   km s$^{-1}$   &   km s$^{-1}$   \\\hline
\noalign{\smallskip}
5013.39  &  -13.792   &   0.058  &  5039.86  &   -14.48  &  0.49  \\
5025.45  &  -14.094   &   0.050  &  5354.78  &    -8.73  &  0.67  \\
5038.44  &  -14.289   &   0.056  &  5405.42  &    -6.59  &  1.40  \\
5041.44  &  -14.285   &   0.038  &  5636.98  &   -12.50  &  0.39  \\
5041.44  &  -14.285   &   0.038  &  5967.19  &   -16.19  &  0.45  \\
5041.44  &  -14.285   &   0.038  &  6042.55  &   -17.76  &  0.46  \\
5052.41  &  -14.002   &   0.061  &  6152.69  &   -15.68  &  0.53  \\
5052.41  &  -14.002   &   0.061  &  6311.77  &   -14.85  &  0.53  \\
5052.41  &  -14.002   &   0.061  &  6416.47  &   -14.43  &  0.46  \\
5341.47  &  -17.593   &   0.088  &  6444.43  &   -11.73  &  0.67  \\
5351.44  &  -17.702   &   0.071  &  6641.25  &    -7.34  &  0.77  \\
5371.45  &  -17.511   &   0.046  &  6698.38  &    -8.50  &  0.45  \\
5414.45  &  -18.018   &   0.055  &  6784.20  &    -9.42  &  0.49  \\
5612.71  &  -15.697   &   0.057  &  6835.45  &   -11.25  &  0.67  \\
5623.65  &  -15.676   &   0.036  &  7105.87  &   -16.53  &  0.49  \\
5638.59  &  -15.469   &   0.050  &  7177.48  &   -15.67  &  0.51  \\
5640.66  &  -15.554   &   0.044  &  7407.78  &   -14.91  &  0.78  \\
5650.61  &  -15.544   &   0.043  &  7473.95  &   -16.71  &  0.58  \\
5655.66  &  -15.489   &   0.051  &                                \\
5941.75  &  -13.256   &   0.034  &                                \\
5945.74  &  -13.637   &   0.108  &                                \\
5965.69  &  -12.998   &   0.151  &                                \\
5966.72  &  -13.035   &   0.033  &                                \\
5991.64  &  -13.084   &   0.067  &                                \\
5991.64  &  -13.084   &   0.067  &                                \\
6027.64  &  -12.950   &   0.039  &                                \\
6042.56  &  -13.003   &   0.035  &                                \\
6057.45  &  -13.260   &   0.046  &                                \\
6137.48  &  -13.088   &   0.030  &                                \\
6140.41  &  -13.214   &   0.040  &                                \\
6145.51  &  -12.953   &   0.047  &                                \\
6187.34  &  -13.345   &   0.035  &                                \\
6304.76  &  -14.073   &   0.041  &                                \\
6318.77  &  -14.238   &   0.023  &                                \\
6393.51  &  -15.236   &   0.038  &                                \\
6439.44  &  -15.640   &   0.041  &                                \\
6444.43  &  -15.912   &   0.037  &                                \\
6617.76  &  -17.490   &   0.046  &                                \\
6664.75  &  -17.677   &   0.048  &                                \\
6669.76  &  -17.695   &   0.040  &                                \\
6708.68  &  -17.357   &   0.038  &                                \\
6716.71  &  -17.253   &   0.034  &                                \\
6762.60  &  -16.795   &   0.062  &                                \\
6792.55  &  -16.425   &   0.051  &                                \\
6797.45  &  -16.709   &   0.027  &                                \\
6835.45  &  -16.132   &   0.035  &                                \\
7083.70  &  -13.805   &   0.063  &                                \\
7098.58  &  -13.664   &   0.050  &                                \\
7109.64  &  -13.605   &   0.045  &                                \\
7131.56  &  -13.682   &   0.054  &                                \\
7165.49  &  -13.327   &   0.031  &                                \\
7189.46  &  -13.303   &   0.030  &                                \\
7407.78  &  -13.110   &   0.035  &                                \\
7465.69  &  -13.372   &   0.044  &                                \\
7466.71  &  -13.196   &   0.053  &                                \\
7489.45  &  -13.598   &   0.050  &                                \\
\hline
\end{tabular}
\end{table}


\begin{table}
\caption{The radial velocities of both components of PG\,1104+243 for the phase binned spectra.}
\label{tb:PG1104_rv}
\centering
\begin{tabular}{lrrrr}
\hline\hline
\noalign{\smallskip}
	&	\multicolumn{2}{r}{MS component}	&	\multicolumn{2}{r}{sdB component}	\\
Phase	&		RV	&	Error		&		RV	&	Error		\\
	&	km s$^{-1}$	&	km s$^{-1}$	&	km s$^{-1}$	&	km s$^{-1}$	\\\hline
\noalign{\smallskip}
0.129	&	-14.99	&	0.042	&	-14.41	&	0.63	\\
0.158	&	-13.963	&	0.046	&	-17.57	&	0.67	\\
0.193	&	-13.078	&	0.047	&	-17.86	&	0.76	\\
0.235	&	-12.403	&	0.027	&	-19.30	&	0.59	\\
0.258	&	-11.970	&	0.026	&	-19.11	&	0.76	\\
0.280	&	-11.568	&	0.026	&	-19.46	&	0.73	\\
0.298	&	-10.925	&	0.036	&	-21.24	&	0.86	\\
0.327	&	-11.423	&	0.027	&	-20.56	&	1.06	\\
0.355	&	-11.243	&	0.025	&	-21.54	&	0.85	\\
0.381	&	-11.234	&	0.022	&	-19.98	&	0.73	\\
0.396	&	-11.323	&	0.030	&	-20.95	&	0.68	\\
0.426	&	-11.563	&	0.023	&	-20.92	&	0.56	\\
0.449	&	-11.699	&	0.056	&	-18.82	&	0.65	\\
0.639	&	-16.109	&	0.054	&	-13.43	&	1.40	\\
0.671	&	-16.988	&	0.047	&	-12.21	&	0.61	\\
0.713	&	-18.018	&	0.024	&	-10.55	&	0.65	\\
0.738	&	-18.570	&	0.021	&	-10.21	&	0.71	\\
0.760	&	-18.967	&	0.024	&	-9.72	&	0.77	\\
0.805	&	-19.681	&	0.027	&	-8.59	&	0.87	\\
0.824	&	-19.774	&	0.025	&	-6.88	&	0.85	\\
0.852	&	-19.966	&	0.025	&	-7.53	&	0.95	\\
0.871	&	-20.089	&	0.037	&	-7.32	&	0.99	\\
0.890	&	-20.085	&	0.021	&	-8.17	&	0.78	\\
0.911	&	-19.735	&	0.022	&	-9.40	&	0.55	\\
0.935	&	-19.486	&	0.061	&	-9.24	&	0.76	\\
\hline
\end{tabular}
\end{table}

\section{SED}\label{a:phot_tables}

\begin{table}
\caption{Photometry of BD-7$^{\circ}$5977 collected from APASS, 2MASS and the Spacelab-1 Very Wide Field Survey of UV-excess objects \citep{Viton1991}.}
\label{tb:BD-7_phot}
\centering
\begin{tabular}{lrrr}
\hline\hline
\noalign{\smallskip}
Band    &   Magnitude    &   Error   \\
        &  mag      &   mag \\\hline
\noalign{\smallskip}
APASS $B$       &  11.061   &   0.031  \\ 
APASS $V$       &  10.506   &   0.082  \\ 
APASS $G$       &  10.801   &   0.081  \\ 
APASS $R$       &  10.298   &   0.094  \\ 
APASS $I$       &  10.125   &   0.037  \\    
STROMGREN $u$   &  11.677   &   0.040  \\ 
STROMGREN $b$   &  10.933   &   0.040  \\ 
STROMGREN $v$   &  11.426   &   0.060  \\ 
STROMGREN $y$   &  10.520   &   0.060  \\
2MASS $J $      &   9.006   &   0.027  \\ 
2MASS $H $      &   8.520   &   0.046  \\ 
2MASS $KS$      &   8.406   &   0.019  \\
\hline
\end{tabular}
\end{table}

\begin{table}
\caption{Photometry of EC\,11031--1348 collected from APASS, 2MASS and UBV photometry from the Edinburgh-Cape Blue Object Survey \citep{Kilkenny1997}.}
\label{tb:EC11031_phot}
\centering
\begin{tabular}{lrrr}
\hline\hline
\noalign{\smallskip}
Band    &   Magnitude    &   Error   \\
        &  mag      &   mag \\\hline
\noalign{\smallskip}
APASS $B$       &  11.061   &   0.031  \\ 
APASS $V$       &  10.506   &   0.082  \\ 
APASS $G$       &  10.801   &   0.081  \\ 
APASS $R$       &  10.298   &   0.094  \\ 
APASS $I$       &  10.125   &   0.037  \\    
JOHNSON $U$     &  11.363   &   0.100  \\
JOHNSON $B$     &  11.840   &   0.100  \\
JOHNSON $V$     &  11.550   &   0.100  \\
2MASS $J $      &   9.006   &   0.027  \\ 
2MASS $H $      &   8.520   &   0.046  \\ 
2MASS $KS$      &   8.406   &   0.019  \\
\hline
\end{tabular}
\end{table}

\begin{table}
\caption{Photometry of TYC\,2084--448--1 collected from APASS and 2MASS.}
\label{tb:TYC2084_phot}
\centering
\begin{tabular}{lrrr}
\hline\hline
\noalign{\smallskip}
Band    &   Magnitude    &   Error   \\
        &  mag      &   mag \\\hline
\noalign{\smallskip}
APASS $B$       &  11.768   &   0.052  \\ 
APASS $V$       &  11.513   &   0.036  \\ 
APASS $G$       &  11.616   &   0.038  \\ 
APASS $R$       &  11.494   &   0.018  \\ 
APASS $I$       &  11.452   &   0.030  \\    
2MASS $J $      &  10.849   &   0.027  \\ 
2MASS $H $      &  10.635   &   0.031  \\ 
2MASS $KS$      &  10.592   &   0.022  \\
\hline
\end{tabular}
\end{table}

\begin{table}
\caption{Photometry of TYC\,3871--835--1 collected from APASS and 2MASS.}
\label{tb:TYC3871_phot}
\centering
\begin{tabular}{lrrr}
\hline\hline
\noalign{\smallskip}
Band    &   Magnitude    &   Error   \\
        &  mag      &   mag \\\hline
\noalign{\smallskip}
APASS $B$       &  11.673   &   0.028  \\ 
APASS $V$       &  11.408   &   0.046  \\ 
APASS $G$       &  11.554   &   0.026  \\ 
APASS $R$       &  11.417   &   0.070  \\ 
APASS $I$       &  11.457   &   0.039  \\    
2MASS $J $      &  10.892   &   0.023  \\ 
2MASS $H $      &  10.698   &   0.023  \\ 
2MASS $KS$      &  10.640   &   0.014  \\
\hline
\end{tabular}
\end{table}

\end{appendix}
\end{document}